\documentstyle[emulateapj]{article}

\begin{document}

\title{Spectral Energy Distributions of Seyfert  
Nuclei\footnote{Based
on observations with the NASA/ESA Hubble
Space Telescope, obtained at the Space Telescope Science Institute, which
is operated by the Association of Universities for Research in Astronomy,
Inc. under NASA contract No. NAS5-26555}}

\author{Almudena Alonso-Herrero\altaffilmark{2}}
\affil{Steward Observatory, The University of Arizona, 933 N. Cherry Ave., 
Tucson, AZ  85721 \\ E-mail: aalonso@as.arizona.edu}
\altaffiltext{2}{Visiting Astronomer at the Infrared
Telescope Facility, which is operated by the
University of Hawaii under contract from the
National Aeronautics and Space Administration.}
\author{Alice C. Quillen\altaffilmark{2}}
\affil{Department of Physics and Astronomy, University of Rochester, 
Rochester, NY 14627}
\author{George H. Rieke}
\affil{Steward Observatory, The University of Arizona, 933 N. Cherry Ave., 
Tucson, AZ  85721}
\author{Valentin D. Ivanov\altaffilmark{2}}
\affil{European 
Southern Observatory, Ave. Alonso de C\'ordova No. 3107, 
Santiago 19001, Chile} 
\and
\author{Andreas Efstathiou}
\affil{Department of Computer Science and Engineering, Cyprus College, 
6 Diogenes street, 1516 Nicosia, Cyprus}

\begin{abstract}
We present nuclear spectral energy distributions  (SEDs)
in the range $0.4-16\,\micron$
for an expanded CfA
sample of Seyfert galaxies.
The spectral indices ($f_\nu \propto \nu^{-\alpha_{\rm IR}}$) 
from $1-16\,\micron$ range from $\alpha_{\rm IR} \sim$ 0.9 to 3.8.   
The shapes of the spectra are correlated with
Seyfert type in the sense that steeper nuclear SEDs ($\nu f_\nu$ increasing
with increasing wavelength) tend
to be found in Seyfert 2s and flatter SEDs 
($\nu f_\nu \simeq$ constant) in Seyfert $1-1.5$s.
The galaxies optically classified as Seyferts 1.8s and 
1.9s display values of $\alpha_{\rm IR}$ as in type 1 objects, or
values intermediate between those of Seyfert 1s and Seyfert 2s. 
The intermediate SEDs of many Seyfert $1.8-1.9$s may be consistent with 
the presence of a pure Seyfert 1 viewed through a moderate amount 
($A_V \lesssim 5\,$mag) of foreground galaxy extinction.
We find, however, that between 10 and 20\% of galaxies with broad optical 
line components have steep infrared SEDs. 

Torus models usually adopt high equatorial opacities  to 
reproduce the infrared properties of Seyfert 1s and 2s, 
resulting in a dichotomy of infrared SEDs (flat for type 1s, and 
steep for type 2s). Such a dichotomy, however, is 
not observed in our sample. 
The wide range of spectral indices observed in the type
2 objects, the lack of extremely steep SEDs, and the large numbers
of objects with intermediate spectral indices cannot be reconciled with predictions
from existing optically thick torus models. We discuss possible modifications 
to improve torus models, including low optical depth tori, clumpy dusty tori, and 
high-optical-depth tori with an extended optically thin component.

\end{abstract}

\keywords{Galaxies: active --- Galaxies: nuclei --- Galaxies: photometry
--- galaxies: Seyfert --- infrared: galaxies}

\section{Introduction}

The discovery that Seyfert 2 galaxies such as NGC~1068
can have reflected or polarized broad-line emission
has led to a 'unified' model in which apparent differences in the properties 
of active galactic nuclei (AGNs) are interpreted in terms of viewing 
angle toward a similar intrinsic underlying structure (Antonucci 1993).
In this unification paradigm, an optically thick circumnuclear dusty torus absorbs a 
significant fraction of the optical/UV/X-ray
luminosity of the active nucleus and reradiates this energy at
infrared wavelengths. One of the challenges of the unified model has been to reconcile
the observed infrared emission with that predicted from the absorbing torus.

Important observational constraints for the torus models come from 
the shape of the infrared spectral energy distributions (SEDs) and the silicate feature at $9.7\,\micron$. 
Previous studies have shown that Seyfert galaxies 
are stronger radiators in the mid-infrared
than most non-Seyfert galaxies (e.g., Spinoglio et al. 1995; Krabbe, 
B\"oker, \& Maiolino 2001)
and that their infrared continua have a range of shapes
from nearly flat ($\nu f_\nu \sim$ constant)
to steeply raising with increasing wavelengths
in the region between 1 and $10\,\micron$. 
Seyfert 2 galaxies generally have SEDs steeper than
Seyfert 1 galaxies (e.g., Rieke 1978; 
Edelson, Malkan, \& Rieke 1987; Ward et al. 1987; Fadda et al. 1998; 
Alonso-Herrero et al. 2001). Torus models predict SEDs that are 
strongly dependent on the orientation of the torus 
with respect to the line of sight of the observer, 
as well as on the density distribution and geometry of the
dusty absorbing material (Pier \& Krolik 1993; Granato \& Danese 1994;  
Efstathiou \& Rowan-Robinson 1995; Granato, Danese,
\& Franceschini 1997). The emission between 
1 and $10\,\micron$ is thought to arise from hot dust 
(up to $\sim 1000\,$K) (e.g., Rieke \& Lebofsky 1981; Barvainis 1987) 
in the inner part of the torus and hence hidden from sight in 
edge-on systems that appear to be of type 2.
To match the $9.7\,\micron$ silicate feature, 
apparently often in absorption in Seyfert 2 and not 
present or slightly in emission in Seyfert 1 spectra
(Rieke \& Low 1975a; Roche et al. 1991; Clavel et al. 2000),
the tori are assumed to be optically thick even to far-infrared wavelengths.

Studies of AGNs
have often focused on high luminosity objects (QSOs and bright Seyfert 1s)
since in these objects the active nucleus
dominates the emission of the host galaxy. These studies avoid the complications
of interpretation when the host galaxy emission is a 
significant fraction of the total (e.g., Kotilainen et al. 1992;
Alonso-Herrero, Ward, \& Kotilainen 
1996).  Because modelling of SEDs is
only available for a handful of relatively high-luminosity 
nearby AGNs (e.g., Efstathiou, Hough, \&
Young 1995; Alexander et al. 1999; Ruiz et al. 2001), the torus parameters are 
not well constrained. The high angular resolution of 
the Hubble Space Telescope ({\it HST}) now 
allows us to probe the nuclei with a beam area about 30
times smaller than is typically achieved with
ground-based observations at visible and near-infrared wavelengths.
Imagers in the mid-infrared also provide a significant improvement in
angular resolution over previous studies.
Such data enable us to separate the nuclear emission from that of
the surrounding galaxy with unprecedented accuracy as well as
to probe for lower luminosity active nuclei.
Recent imaging from the Infrared Space Observatory  ({\it ISO})
also allows us to measure the nuclear emission in the mid-infrared
with improved sensitivity compared with previous observations.

We have used these capabilities on an expanded version of the 
CfA sample of Seyfert galaxies. 
The infrared SEDs of the CfA sample have been studied 
by Edelson et al. (1987), Pier \& Krolik
(1993), Maiolino et al. (1995) and P\'erez Garc\'{\i}a \&
Rodr\'{\i}guez Espinosa (2001).  Fadda et al. 
(1998) compiled data from the literature and {\it IRAS} to produce
infrared SEDs of a heterogeneous sample of Seyfert 
galaxies, but they mainly focused their efforts
on fitting infrared colors to their model outputs. All these works 
noted that the relatively large apertures used were likely to be contaminated
by emission from the host galaxies, a particular problem for
the Seyfert 2 galaxies.  
To address this concern, in this paper we present high resolution IRTF
$K^\prime$ ($2.1\,\mu$m) and $L$ ($3.5\,\mu$m) band
imaging of the CfA Seyfert galaxies, from which we have estimated 
the non-stellar nuclear
fluxes. We combine these fluxes with measurements of the unresolved
nuclear emission from optical {\it HST}/WFPC1 and 
{\it HST}/WFPC2 and near-infrared 
{\it HST}/NICMOS and mid-infrared measurements
made from ISOCAM images, and/or small-aperture ground-based 
$10\,\mu$m measurements from the literature. This suite of data is used 
to compare
the properties of the non-stellar SEDs with the predictions of 
models for dusty tori.

\section{The sample}

\subsection{The CfA sample}
Tables~1 and 2 (see also Section~2.2) 
list the sample of galaxies on which this paper is based. 
The galaxies 
are from the CfA redshift survey, which is drawn from the fraction of the sky
defined either by
$\delta \ge 0^\circ$   and $b \ge 40^\circ$ or
$\delta\ge -2^\circ.5$ and $b \le-30^\circ$, and $m_{\rm Zw}
\leq 14.5$.
The AGNs were identified by means of optical spectroscopy, thus defining
a ``complete'' magnitude-limited sample of 49 Seyfert
galaxies (Huchra \& Burg 1992). Because it is not color-selected, this sample should
be relatively free of selection effects that tend to enhance
the proportion of galaxies with anomalously strong emission in the color
used for selection (Huchra \& Burg 1992; Osterbrock \& Martel 1993). 

Spectroscopic classifications for the CfA survey 
are taken from Osterbrock \& Martel (1993) and references therein. However
Ho, Filippenko, \& Sargent (1997) identified 5 Seyfert 2
galaxies as having lower Seyfert
types than listed by Osterbrock \& Martel (1993).  In these cases
we list in Table~1, column~2 
the identification from Osterbrock \& Martel (1993) on the left
and that from Ho et al. (1997) on the right.
Ruiz, Rieke, \&  Schmidt  (1994) detected broad components  in near-infrared 
lines in Mkn~334 and NGC~5252 (He\,{\sc i} at $1.083\,\mu$m) 
and Mkn~533/NGC~7674 (He\,{\sc i} and Pa$\beta$). 
 Optical polarized broad-line components were detected by Young et al. (1996)
in NGC~4388, NGC~5252 and Mkn~533.
Tran (2001) has found a hidden broad line region in 
NGC~7682. For the Seyfert 2s in Table~1 we also indicate whether 
such hidden broad-line regions have been detected.

\subsection{The extended CfA sample of Seyfert galaxies and its 
completeness}

The CfA sample is usually regarded as a complete
sample of optically selected 
Seyfert galaxies. However Maiolino \& Rieke (1995) show 
that it is incomplete both in terms of low luminosity nuclei 
and edge-on galaxies. Ho \& Ulvestad (2001) also discuss selection biases that may affect 
the completeness of CfA sample. Because of 
the method used to identify AGNs in the CfA sample, only 
Seyfert 2s having the broadest, most extended wings in the observed
emission lines were included in the final sample. The distinction 
in the original classifications between LINERs and Seyferts is also not rigorous. According to
these authors the CfA sample is likely to be complete only for
relatively bright objects, and the Seyfert 2s in particular are 
intrinsically more luminous and more radio-powerful than is typical. 

To alleviate in part the possible biases in the original CfA sample, we have
defined an extended sample. It includes all the Seyfert galaxies in the original CfA sample as well as 
those galaxies initially classified as LINERs by Huchra \& Burg (1992), but later reclassified as 
Seyfert galaxies by Ho et al. (1997) from high S/N optical 
spectra. We therefore add a total of nine galaxies (see Table~2) to the original CfA sample, 
thus increasing its size by $\simeq 20\%$ (a total of 58 galaxies).

Our sample is still probably weighted against low-luminosity type 2 Seyferts. 
Table~3 compares the fractions of type 1 ($1-1.5$) and 
type 2 ($1.8-2$), as classified from 
optical spectroscopy in a number of nearby AGN samples: 1.) the RSA 
sample (see  Maiolino \& Rieke 1995); 2.) the Palomar survey (Ho et al. 1997);
and 3.) the infrared-selected 
$12\,\micron$ sample (Spinoglio \& Malkan 1989). Since the CfA sample 
shows a deficit of highly inclined galaxies (see Maiolino 
\& Rieke 1995, and Section~5.5), in this table 
we make the same comparison for galaxies with axial ratios 
$b/a \ge 0.6$. Galaxy inclinations are taken from Ho et al. (1997) and from the RC3 
(de Vaucouleurs et al. 1991). We find that in low inclination galaxies, 
both the extended CfA sample and 
the $12\,\micron$ sample have slightly smaller fractions of type 2s 
than the Palomar and RSA samples, although the numbers are not different
at a statistically significant level. 
If, as claimed by Ho \& Ulvestad (2001), the CfA sample only contains
Seyfert 2s with relatively bright broad emission lines, then the CfA 
sample would be biased  against low luminosity type 2 AGNs rather than against 
pure type 2 objects (that is, those with an equatorial or near equatorial 
view of the AGN). And indeed, it appears from Table~3 that the Palomar survey has the highest
fraction of type 2 objects, probably because the high quality of its 
spectra allows detection of very low luminosity objects. 

\section{Observations}

\subsection{IRTF $K^\prime$ and $L$ band observations}
We have obtained high resolution $K^\prime$ ($\lambda_{\rm central} =
2.12\,\mu$m) and $L$ band ($\lambda_{\rm central} =
3.51\,\mu$m) imaging of 
34 CfA Seyfert galaxies using NSFCam (Rayner et al. 1993; Shure et al. 1994)
at the 3.8\,m NASA IRTF telescope on Mauna Kea. The observations were made
in August 1999, March 2000, August 2000 and May 2001.
We used the 0.3\arcsec/pixel plate scale. The nuclei of the galaxies 
were sufficiently bright that we were able to 
use the tip-tilt capability during  the first three campaigns (it was not available 
during the fourth campaign), thus improving the image quality.  
Typical integration times were 300 seconds in the $K^\prime$
band and $720-1200$ seconds (depending on the
galaxy brightness) in the $L$ band. In Figure~1 we show three galaxies 
displaying extended emission in the $L$ band. The spatial resolution (FWHM) 
in the $L$ band 
as measured from the standard stars  was $0.6\arcsec-0.9\arcsec$.

Standard  data reduction procedures were applied for the $K^\prime$ 
images and for bright sources 
in the $L$ band.  Because of the short individual exposures in the 
$L$ band (to avoid saturation), a number of galaxies were not detected on individual 
exposures, and thus the images had to be 
shifted to a common position and  coadded using the nominal telescope offsets (blind offsets,
see Table~4). In these cases the morphology and flux of the 
unresolved component cannot be determined, 
and only the integrated flux can be measured.

Conditions were photometric during all runs except for one night during the  
August 2000 campaign. Throughout the nights we obtained 
observations of standard stars from Elias et al. (1982) for 
photometric calibration. The typical rms scatter in these calibration observations is 
$0.04-0.06\,$mag in both $K^\prime$ and $L$ bands. The major source of uncertainty is 
the background subtraction, especially 
at the longer wavelengths ($L$) for the faint sources. In Table~4 we give 
the K$^\prime$ and L band aperture photometry.

\subsection{Optical and near-infrared wavelengths: {\it HST} archival images}

{\it HST}/NICMOS images of the CfA sample 
were taken for 22 Seyfert $1.8-2$ galaxies 
with the NIC1 camera (0.043\arcsec/pixel) 
at $1.1\,\micron$ (F110W filter) and $1.6\,\micron$ (F160W filter)
as part of proposal 7867 and are described by Martini \& Pogge (1999)
and Martini et al.  (2001). NGC~1068 was a guaranteed time target.   
In addition, some of the CfA Seyfert $1-1.5$ galaxies were observed as 
part of a number of GTO and GO proposals, mainly through the 
NIC2 (0.076\arcsec/pixel) 
F160W filter. The F110W and F160W filters are roughly equivalent 
to the ground-based broadband $J$ and $H$ filters. Details on the 
{\it HST}/NICMOS observations, data reduction and 
fluxes for the CfA Seyferts can be found in Quillen et al. (2001).

When possible we also used
WFPC2 archival images through the F606W filter 
(described by Malkan, Gorjian, \& Tam 1998). If these images were 
saturated we used images taken through other filters or  
obtained the fluxes for the 
unresolved component from the literature (Ho \& Peng 2001 for WFPC1 or WFPC2).  
See Tables~1 and 2 for the filters.

\subsection{Mid-infrared wavelengths}
Mid-infrared (6.75 and $9.63\,\mu$m) fluxes measured from
{\it ISO}/ISOCAM images were taken from Clavel et al. (2000). 
The FWHM of the ISOCAM PSF is approximately 
$\sim 4-5\arcsec$ (Clavel et al. 2000).  
For a few galaxies which were not part of this program 
we obtained ISOCAM reduced images from the {\it ISO} archive.
We also make use of the $16\,\mu$m {\it ISO}/ISOPHOT 
fluxes measured for the CfA Seyferts by P\'erez Garc\'{\i}a \& 
Rodr\'{\i}guez Espinosa (2001). In addition, we obtained 
ground-based small aperture $N$-band ($10.6\,\micron$) fluxes 
from the literature, both to compare with the lower resolution
ISOCAM observations, and to use for those galaxies where
ISOCAM data are not available. 

In Table~1 and Table~2 we summarize the available 
{\it HST}/WFPC1, WFPC2 and NICMOS, IRTF $K^\prime L$ 
and {\it ISO} and ground-based mid-infrared observations 
for the CfA Seyfert galaxies, along with appropriate references when the 
data were taken from the literature.

\section{The unresolved emission}
The contribution from the underlying galaxy can 
dominate the central emission in Seyfert 2 galaxies, especially at 
wavelengths shorter than approximately $2\,\mu$m (see e.g., Alonso-Herrero,
Ward, \& Kotilainen 1996; Alonso-Herrero et al. 
2001) and can also be  significant in Seyfert 1 galaxies 
(e.g., Kotilainen et al. 1992). It is therefore essential to eliminate 
the stellar contributions to isolate the non-stellar emission from the AGN.
Since the data used in this work have been obtained with 
different instruments, the methods for obtaining the unresolved emission 
are slightly different, as we describe below.

\subsection{HST images}

In Quillen et al. (2001) we estimated the flux from the unresolved 
component in the F160W filter ($1.6\,\micron$) by fitting the sum of a
galaxy profile and the point spread function (PSF) for a large
number of Seyfert galaxies. 
For  the underlying galaxy the profiles were fitted by 
exponential and power laws convolved with the appropriate NICMOS
PSF generated by Tinytim (Krist et al. 1998). Here we applied
the same procedure for the NICMOS NIC1 F110W images ($1.1\,\micron$) 
and the WFPC2 optical images. In addition we obtained nuclear 
WFPC1 or WFPC2 fluxes 
for a number of galaxies from Peng \& Ho (2001), see Tables~1 and 2. 
These authors fitted 
a two dimensional model consisting of an analytical description 
for the bulge light plus an additional point source for the nucleus, convolved
with an appropriate synthetic PSF. Their modeling for the unresolved 
component is thus similar to ours.

\subsection{$K^\prime$ and $L$-band images}

We estimated the non-stellar fractions at $K^\prime$ and $L$ band 
by fitting the PSF surface brightness profile to the nucleus of the galaxy, and assumed that the 
fitted PSF integrated flux represents the unresolved
component (see Alonso-Herrero et al. 1998 for
more details). The FWHMs of the PSF in the K$^\prime$ and 
$L$ band were measured from standard star observations
and from stars present in the images (only in $K^\prime$). 
We also produced surface brightness profiles of the host galaxies to determine the 
amount of extended emission.

In Table~4, column~(2)
gives the measured FWHM of the nuclear emission 
in the $L$ band, column~(6) is the fraction of unresolved 
emission in a 3\arcsec-diameter aperture, and  column~(7) is the nuclear 
morphology in the $L$ band. As can be seen from this table
the fraction of unresolved emission in 
the $L$ band within a 3\arcsec-diameter aperture 
varies from $16$ to $100\%$, although
in most cases is within $40-70\%$. McLeod \& Rieke (1995) also
obtained estimates of the unresolved fluxes in the $K$ band. We find 
that our estimates agree with theirs to within the photometric and 
fitting errors of both works.

\subsection{Mid-infrared wavelengths and contribution from 
star formation to ISO fluxes}

In AGNs past $\simeq 3\,\micron$ we expect the contribution 
from the underlying old stellar population to the nuclear
fluxes to be much reduced compared
to that at shorter wavelengths. However if the nuclear
stellar population is dominated by a starburst 
there may be contamination at wavelengths $> 3\,\mu$m.
We note that recently Krabbe et al. 
(2001) and Lira et al. (2001) have obtained ground-based $N$ imaging
of small samples of Seyfert galaxies and found that the mid-infrared 
$10\,\micron$ emission is dominated by
a central unresolved source; however, a low surface brightness extended
component powered by star formation might not be apparent in such data. 

Because of the relatively large {\it ISO}/ISOCAM 
aperture (diameter of 18\arcsec, see Clavel
et al. 2000) we need to determine whether the 
$6.75$ and $9.63\,\micron$ {\it ISO} fluxes are 
contaminated by circumnuclear star formation. To this end we
compare 6\,cm radio measurements through a 
large ($\simeq 15\arcsec$) aperture with 
those through a small ($\simeq 0.3\arcsec$) aperture. The radio 
measurements are taken from Kukula et al. (1995) 
for the small aperture and from Rush, Malkan, \& Edelson (1996) for 
the large aperture. The small aperture 
radio measurements of Seyfert galaxies are most likely  to originate in the 
AGN whereas the large aperture fluxes will include contributions from 
both the AGN and extended star formation. Small 6\,cm 
$f_{0.3\arcsec}/{f_{15\arcsec}}$ ratios are therefore a 
warning that there may be a significant level of 
circumnuclear star formation within the large aperture.  

In Table~5 we list Seyfert galaxies in the CfA 
sample with available {\it ISO} fluxes and fractions of nuclear 
to 15\arcsec \ radio emission of
$\le 15\%$. For comparison in this table we also list, when 
available, the $10\,\micron$ ground-based\footnote{Typically the 
apertures of ground-based data were of the order of $5$-$8.5\arcsec$.}  
to {\it ISO} flux ratios, and whether there is evidence for 
extended emission in the $L$ band and {\it ISO} images. The latter 
are from Clavel et al. (2000) and our own analysis of the images.
We have also included in this table NGC~5929 and NGC~5940 as they
appear to have extended mid-infrared emission from the comparison 
of the {\it ISO} and ground-based $10\,\micron$ fluxes.

Mkn~334, Mkn~841 and NGC~4051 have small 
$f_{0.3\arcsec}/{f_{15\arcsec}}$ ratios at 6\,cm, but do not appear to 
be extended in either the {\it ISO} or the $L$-band images, and 
have mid-infrared ground to {\it ISO} ratios consistent with 
most of this emission stemming from a small region. 
NGC~4388 shows evidence for extended radio emission, but the   
{\it ISO} images reveal the  presence of a bright point source 
on top of an extended diffuse component, so it is likely that 
most of the {\it ISO} fluxes are unresolved emission originating in the AGN. 

NGC~1144, NGC~3982, NGC~5033 and Mkn~266SW 
are cases where the {\it ISO} mid-infrared fluxes
may contain a significant contribution from circumnuclear star formation. 
In NGC~1144 there are H\,{\sc ii} regions a few arcseconds SW
of the nucleus (see Figure~1 and also 
Hippelein 1989). In NGC~3982 there are H\,{\sc ii}
regions within the 18\arcsec \ {\it ISO} aperture 
(Gonz\'alez-Delgado et al. 1997), whereas in NGC~5033 bright H\,{\sc ii}
regions are present in the spiral arms to within 5\arcsec \ of 
the nucleus (Evans et al. 1996). Mkn~266SW is part of an interacting
system whose nuclei are separated by $\simeq 10\arcsec$. Since
we detect $L$-band emission from both nuclei (see Table~4) it is 
possible that the {\it ISO} fluxes also contain contributions from both 
nuclei. For these galaxies we list in Table~6 small aperture 
ground-based mid-infrared fluxes.
If these are not available then the {\it ISO} fluxes
are given as upper limits to the unresolved emission.
Although there may still be some contribution from a compact starburst
even for the small aperture ground-based measurements, it will
be energetically less important than that from extended star formation 
(Imanishi 2002).

\section{Observed Spectral Energy Distributions}

In Tables~6 and 7, columns~(2) through (9), 
we have summarized the optical and near- and 
mid-infrared unresolved (non-stellar) fluxes 
for the CfA Seyfert 2s and Seyfert 1s, respectively. Errors in all the
unresolved fluxes 
are on average $\pm 30\%$. We also indicate in these two tables, when 
appropriate, the upper limits to 
the nuclear fluxes. Because of the large aperture of 
the {\it ISO}/ISOPHOT $16\,\mu$m measurements, we 
show all the fluxes at this wavelength as upper limits as they 
may be contaminated to some 
degree by emission from the host galaxy.

In Figure~2 we present the non-stellar SEDs of NGC~1068 and 
NGC~4151, where the 
1.1 through $2.2\,\mu$m data points are from 
{\it HST}/NICMOS, and the 3.8 and 
$4.8\,\mu$m are ground-based data, all taken 
from Alonso-Herrero et al. (2001) and references therein. 
We also show the {\it ISO} spectrum 
of the AGN component of NGC~1068 (Le Floc'h et al. 2001), as well 
as the {\it ISO}/ISOPHOT $16\,\mu$m data points  from P\'erez Garc\'{\i}a \& 
Rodr\'{\i}guez Espinosa (2001). The additional mid-infrared
 wavelength points are
ground-based observations through small ($6 - 8.5\arcsec$) apertures
(Rieke \& Low 1972; 1975a; 1975b) and diffraction limited 
measurements of the central emission
(Alloin et al. 2000 for NGC~1068 and Radomski et 
al. 2002 for NGC~4151). These two galaxies provide a guide to the behavior of
the other galaxies in the sample, for which less complete data
are available. From 10 to 34$\mu$m, the SEDs are very similar, and
both are flat in $\nu f_\nu$. From 1 to 10$\mu$m, the SED
of NGC~1068 has the appearance of a very strongly reddened version
of the SED of NGC~4151. Qualitatively, this difference is as
predicted by the unified model. 

The non-stellar optical through mid-infrared 
($\simeq 16\,\micron$) SEDs of the remaining galaxies are presented 
in Figures~3, 4  and 5 for the CfA Seyfert 
$1-1.5$s, Seyfert $1.8-1.9$s and Seyfert 2s, respectively.
When plotting the SEDs and fitting the spectral 
indices (next section) the ground-based $N$-band fluxes
(derived to give monochromatic fluxes of Rayleigh-Jeans
spectra) need to be corrected to $\nu \ f_{\nu} = {\rm constant}$ by 
multiplying them by a factor of 1.22 (see also Edelson et al. 1987).

Variability in the near-infrared and optical
region (e.g., Clavel, Wamsteker, \& Glass 1989) is a concern
when assembling SEDs of AGN from multiple data sources.   
Most of these data were observed within
a time period of a couple years: the {\it ISO} lifetime was Nov.~1995 through May~1998,
and the NICMOS lifetime prior the cryocooler era was Feb.~1997 through Dec.~1998.
Most of the WFPC2 images were observed within a year of these operational lifetimes.
A number of the CfA galaxies 
have duplicate NICMOS observations. NGC~5273, NGC~5033, NGC~5347, UGC~12138,
and UM~146 displayed variable levels (over a time scale of
a few months) of the unresolved emission at $1.6\micron$
but Mkn~573, NGC~5252 and Mkn~471 did not (Quillen et al. 2000).
Typical levels of variations in the near-infrared (at $1.6\,\micron$) 
fluxes are up to $\sim 20\%$ (Quillen  et al. 2000).  
This level of variation should be  within the uncertainties of the 
unresolved flux estimates in the near-infrared,  
 so we do not expect that our SEDs are strongly
affected by intrinsic variability.

A common parametrization to describe the non-stellar 
SEDs of AGN is to fit a spectral  
index ($\alpha$) so that the observed fluxes can be expressed as power laws,  
$f_\nu \propto \nu^{-\alpha}$. For a significant fraction of the Seyfert 
2s we only have upper limits to the unresolved emission -- 
at short wavelengths if a point 
source was not detected, and at longer wavelengths if
there is evidence for stellar emission. 
We use the package {\sc asurv} (La~Valley, 
Isobe, \& Feigelson 1992) to fit 
the SEDs because it allows us to 
include upper limits.
We have fit spectral indices in the optical to infrared 
(up to $16\,\micron$) spectral range
($\alpha_{\rm opt-IR}$) and in the infrared ($1-16\,\micron$) 
spectral range alone
($\alpha_{\rm IR}$). The spectral
indices are presented
in the last two columns of Tables~6 and 7. In Figure~6 we show 
the distributions of the fitted spectral indices in terms
of the Seyfert type as derived from optical spectroscopy.

\subsection{The effects of extended foreground extinction in the host galaxy}

Before we discuss the observed SEDs of the CfA Seyferts and compare them with 
torus models, we must consider the effects of dust in the host galaxy.
Most of the Seyfert galaxies in the 
CfA sample have dust near their nuclei
(Martini \& Pogge 1999; Pogge \& Martini 2002).  Generally the extinction
is not strong enough to be prominent in the NICMOS
images (see Martini \& Pogge 1999), which sets a limit on the total
extinction that can be present on scales resolvable by {\it HST}.
As can be seen from their color maps, the dust is generally located in 
front of the nucleus (see also Regan \& Mulchaey 1999).
We can therefore consider it as a purely foreground medium.
Martini \& Pogge (1999) used the $V-H$ colors of the Seyfert 2s in 
the CfA sample to derive moderate levels of optical extinction, 
up to $A_V \sim 5\,$mag. 

To illustrate the effects of dust in the host galaxy, 
we took SEDs produced by the tapered torus+cone models (see
Section~6 and Table~8) from
Efstathiou et al. (1995) and inferred their appearance as
seen through a foreground dust screen.  We consider three different 
viewing angles to the AGN: $\theta_{\rm v} =0\arcdeg$ 
(polar view of the AGN), $\theta_{\rm v} = 30\arcdeg$ (intermediate
view), and $\theta_{\rm v}= 90\arcdeg$ (equatorial, through the torus, 
view of the AGN). The intermediate viewing angle has been chosen to 
coincide with the half opening angle of this particular torus model, which 
corresponds to the transition between type 1 and type 2 objects. 
For the foreground extinction we used values of $A_V = 0.4, \  
1 \ {\rm and} \  5\,$mag, and the Rieke \& 
Lebofsky (1985) extinction curve. The resulting
spectra are shown in Figure~7. The reddened spectrum from a nearly face-on (polar view) torus 
(underlying Seyfert 1, $\theta_{\rm v}=0\arcdeg$) 
will look similar to the intrinsic spectrum at a 
viewing angle closer to a true Seyfert 2 ($\theta_{\rm v} = 90\arcdeg$) 
given sufficient levels of foreground extinction ($A_V > 5$\,mag). 
However, moderate amounts of extinction will not have a significant 
effect on the nuclear fluxes at wavelengths greater than 
approximately $2\,\micron$. In addition, if a foreground screen of dust
that is generally thick enough to modify the infrared SEDs has small holes,
the view of the torus through these holes will dominate the flux
in the near infrared and give an impression of a low level of extinction
overall.

The torus models place the source of the near-infrared emission close
to the broad-line region, and of a diameter of order 1 pc (compared with
the size of about 0.01 pc for the BLR). This general geometry is supported 
by observations of near-infrared variability with timescales of order a 
year (e.g., Rieke \& Lebofsky 1979; Lebofsky \& Rieke 1980; Glass 1998 
and references therein). To obscure the entire near-infrared-emitting 
region with a foreground cloud that nonetheless lets the broad emission 
lines through would take a highly contrived configuration for the 
obscuring cloud: a small hole aligned on the
BLR with no other holes elsewhere where near-infrared light could escape in our direction. 
Thus, it is improbable that dust in the host galaxy could significantly modify the
infrared SED and still allow broad lines to be detected in optical spectra.

\subsection{Seyfert $1-1.5$s}

The most remarkable aspect of the optical to mid-infrared non-stellar  
SEDs for the type $1-1.5$ galaxies (see Figure~3) is the high
degree of uniformity, with a featureless continuum that is
nearly flat in $\nu f_\nu$, similar to that of NGC~4151. Overall, the newly derived SEDs 
are not dramatically different from those inferred in previous works, since
the dominance of the AGN emission over the stellar  
emission in the nuclear regions makes detailed image deconvolution 
unimportant to derive true nuclear fluxes. However if we compare our SEDs on 
a case-by-case basis with those in Edelson et al. (1987), it is clear that their Seyfert 1 SEDs
between $1$ and $16\,\micron$ tend to appear bluer in $\nu\,f_{\nu}$ 
than ours. We believe the fluxes at shorter wavelengths in Edelson et al. (1987) 
may still contain some contribution from stellar emission (these
authors used 8.5\arcsec-diameter apertures for their photometry).

The average values for the optical-infrared and infrared 
spectral indices of the Seyfert $1-1.5$s (22 galaxies)
are $\alpha_{\rm opt-IR}
= 1.59 \pm 0.30$ and $\alpha_{\rm IR} = 1.48 \pm 0.30$, where we have 
excluded NGC~6104 and Mkn~993 (optically classified 
as Seyfert $1.5-1.8$ and Seyfert $1.5-2$, respectively) 
as they clearly show SEDs more compatible with
those of type 2 objects.  For comparison Edelson et al. (1987) 
found $\alpha= 1.15 \pm 0.29$ between $2.2\,\micron$ and $25\,\micron$ for
the CfA sample, and Fadda et al. (1998) measured a median 
spectral index  between $1\,\micron$ and $20\,\micron$ 
of $\alpha= 1.4 $ for a heterogeneous sample of Seyfert 1s. 
The fact that the infrared spectral indices are 
on average bluer than the optical-infrared indices is probably 
due to the presence of some amount of foreground dust in Seyfert 1s.

However, two type 1 galaxies, Mkn~590 and NGC~7603, have 
well-measured SEDs that depart significantly 
from the pattern established for the other type 1 objects. The SEDs
of both start to drop from the near infrared towards the visible,
relative to the flat SEDs (in $\nu f_\nu$) that are typical. 
We quantify this behavior from the nuclear fluxes measured at
$K'$ and $L$ since they use identical observational techniques and
were obtained simultaneously (hence variability is not an issue). 
Relative to the median $K' - L$ for the type 1 nuclei, an extinction
of order A$_V = 10\,$mag is indicated for both galaxies. A similar 
extinction level would apply to the BLR, which would make the
identification of these galaxies as type 1 very improbable. 
We conclude that it is implausible that the differences in the SEDs 
for these two galaxies compared with the other type 1 objects can be explained
in terms of identical intrinsic SEDs modified by foreground extinction.

\subsection{Seyferts $2$s}

In clear contrast with the CfA Seyfert $1-1.5$s, the Seyfert 2s 
show a variety of spectral shapes (see 
Figure~5, and also Figure~6). 
All the Seyfert 2 galaxies have SEDs that rise steeply at increasing 
wavelengths ($\alpha_{\rm IR} > 2.6$), qualitatively similar to the
SED of NGC~1068. This behavior is roughly consistent 
with the unified model predictions for a Seyfert 2 nucleus, that is, a heavily obscured 
version of the typical Seyfert 1 SED. 

The Seyfert 2 nuclei in the CfA sample have near-infrared colors and  
near-to-mid infrared colors in general 
redder than seen in the compilation 
by Fadda et al. (1998), but similar to the Seyfert 2s 
observed by Alonso-Herrero et al. (2001).
One possibility is that the Seyfert
2 photometry compiled by 
Fadda et al. (1998) was still contaminated with light from the galaxy
for the short wavelengths, 
as the determination of the unresolved component is 
usually limited by the seeing conditions when using  
ground-based observations (see Alonso-Herrero et al. 1996).
The infrared spectral index of the median 
SED for a type 2 object in Fadda et al. (1998) is 
approximately $\alpha_{\rm IR} = 2.2$ between 1 and $20\,\micron$. 
However from Figures~5 and 6 it is clear that  
it is not appropriate to define a 'median' SED for type 2 objects, as it is
precisely the variety of spectral shapes that needs to be reproduced 
by the torus models,  as we will discuss in Section~7.

Some of the type 2 nuclei have substantial amounts of extended foreground
dust associated with their host galaxies.  
For instance NGC~4388 and NGC~5929 have measured NLR extinctions 
of $A_V\simeq 4\,$mag (see Veilleux et al. 1997 and Rhee \& Larkin 2002, 
respectively). The effect of correcting the observed SEDs for the
indicated amounts of foreground 
extinction is small. Foreground dust at the level seen in these galaxies can strongly
affect the optical classification of the nuclear activity if it preferentially obscures the
broad-line region (as proposed by Maiolino \& Rieke 1995). However, this
amount of dust does not seem capable of influencing the infrared SEDs sufficiently to
account for the systematic differences seen between Seyfert 1 and Seyfert 2 
SEDs.

\subsection{Seyfert $1.8-1.9$s}

UGC~12138, NGC~3786, NGC~5033, and NGC~5273 (Figure~4) are optically classified  as  
Seyfert $1.8-1.9$s, but have nearly flat SEDs, similar to those of 
Seyfert $1-1.5$s (see Figure~3). In fact their infrared spectral indices
fall well within the values found for the CfA Seyfert $1-1.5$s.
The extinction maps of Martini \& Pogge (1999), based on $V - H$ colors,
show that all four of these galaxies are free of heavy, extended obscuring
foreground clouds in their host galaxies.  

We exclude from classification as type 1.9 those galaxies 
where the broad line component is only seen in the
infrared, e.g., Pa$\beta$. The intermediate-class 
galaxies UM~146, Mkn~334, Mkn~471, NGC~4258, 
NGC~5252, and NGC~5674 (Figure~4), display intermediate
SEDs (infrared spectral indices in the 
range $\alpha_{\rm IR} \simeq 1.8-2.6$) that do not drop toward short
wavelengths as steeply as NGC~1068. In most of these galaxies, unresolved 
nuclear emission is  measured from optical and/or near-infrared {\it HST} images. In
all of these cases, the $V - H$ images of Martini \& Pogge (1999) suggest
a possibility of extended obscuration from the host galaxy, although the
bright and blue nucleus of Mkn~471 indicates that it is probably clear of the nearby clouds.  

One possibility for the Sy$1.8-1.9$ intermediate SEDs is that 
they are those of a true Seyfert 1 galaxy seen 
through some foreground dust. However, an explanation of the steep infrared SEDs in terms of
foreground dust is unlikely in the case of NGC~4258. 
We have used the same approach described under type 1 objects, estimating
the extinction needed to redden a median type 1 infrared SED to agree with
the $K' - L$ colors, and find for this galaxy a level equivalent
to A$_V$ $\ge$ 10, making observation of any broad-line component very
problematic. A similar conclusion holds for NGC~7479; its $K' - L$ is so
red that the nucleus is not detected at the shorter wavelengths, but the
NICMOS upper limits demonstrate that it is heavily obscured (if the intrinsic
SED is Seyfert-1-like). 

We therefore find four cases among the type 1 and intermediate-class Seyfert
galaxies where broad line components are seen despite an infrared SED that
is either intrinsically red or is heavily reddened. The conclusion that broad
line components can escape from systems with such SEDs
is reinforced by the detection by Ho et al. (1997) of broad wings on
H$\alpha$ in NGC~1068, NGC~3982, and NGC~4388.  
For many of this total of seven objects, under the unified
model the "extinction" we see in the infrared SED is likely to be
a property of the torus itself, rather than the result of a foreground obscuring cloud. 

The Seyferts with intermediate SEDs are the most difficult for the
disk and torus models to explain.  The models
are extremely sensitive to the viewing angle.  Only for a very small
range of orientation angles, just grazing the surface of the
disk or torus, are intermediate SEDs predicted. The models would also
predict that the broad emission lines would be obscured at virtually all viewing
angles where the near-infrared emission region is obscured. Yet we find that
$10 - 20\%$ of the galaxies with detectable broad lines in the optical
have red infrared SEDs. That is, a strict dichotomy
of SEDs correlated with emission-line properties is predicted,
but is not in accordance with our measurements.

\subsection{Spectral indices and the inclination of the host galaxy} 

There appears to be a deficit of 
Seyfert $1-1.5$ galaxies in edge-on galaxies (e.g., 
Keel 1980; Lawrence \& Elvis 1982; Maiolino \& Rieke 1995; 
Schmitt et al. 2001) in most of the optically selected 
samples. In particular, Maiolino \& Rieke (1995) concluded that the
CfA sample has a deficiency of Seyfert 1s in edge-on galaxies, caused by an extended
torus (scale of a 100\,pc)  coplanar with the host 
galaxy that obscures the BLR. They predict that the spectrum of a Seyfert 1 galaxy will 
look Seyfert 2-like in many of these cases. 

The behavior of the infrared SEDs provides an independent way to
estimate whether AGNs are of type 1 or 2. Where the nuclei may be
obscured by dust in the host galaxy, the SEDs are much less affected 
than the optical emission lines. Using the SED shapes, we can check 
whether the portion of intrinsically type 1 galaxies is similar in
edge-on systems to the portion found in optically selected samples in
face-on galaxies.

The axial ratios ($b/a$) for the 
extended CfA sample of Seyfert galaxies are taken from 
McLeod \& Rieke (1995) and Ho et al. (1997). In Figure~8 we show
the spectral indices as a function of the host galaxy inclination, $b/a$. 
The galaxies have been divided in Seyfert 1s (1 and 1.5 types) 
and Seyfert 2s (1.8, 1.9 and 
2 types) according to their optical classification. We also mark 
in this figure for reference an axial ratio of $b/a=0.6$ (or inclination of 
the host galaxy of $i=53\arcdeg$). Maiolino \& Rieke (1995) found 
that pure Seyfert 1 galaxies with axial ratios of 
$b/a<0.6$ will be most likely classified as
Seyfert 1.8 or 1.9. This result is confirmed by Figure~8.

If we use the infrared SED to divide galaxies into probable 
type 1 and type 2, we can see that 
in terms of the spectral index $\alpha_{\rm IR}$ there 
are approximately equal numbers of
type 1 ($\alpha_{\rm IR} = 1.52 \pm 0.38$, 6 galaxies) and type 2 
(5 galaxies) in the highly inclined galaxies in 
the extended CfA sample (Figure~8, right panel). These 
numbers are consistent those for the face-on members of the sample. 
This result confirms the suggestion by Maiolino \& Rieke (1995) 
that Seyfert 1 nuclei hosted in 
highly inclined galaxies may be misclassified as type 2 objects. 
For example, studies based on optical spectroscopy and that ignore 
inclination effects will overestimate the portion of type 2 nuclei.

\section{Model predictions}

For the following discussion, we assume the validity of a unified model,
that is we assume that all Seyfert galaxies have similar dusty tori
in their nuclei. We will try to find a geometry for these
tori that is consistent with our observations.

The broadness of the spectrum in the $1-20\,\micron$ region
and the nature of the $9.7\,\micron$ silicate feature (in absorption and 
emission) have been a challenge for the torus models.  
Compact torus models (Pier \& Krolik 1993) often predict spectra that are not 
sufficiently broad (they lack emission in the near-infrared)
or they emit too strongly in the $9.7\,\micron$ silicate feature for
polar orientations.
This issue has resulted in a suite of more extended disk models:
the flared disk (Granato et al. 1997), 
the tapered disk (Efstathiou \& Rowan-Robinson 1995), and the warped disk 
(Sanders et al. 1989).
Excess emission in the near-infrared in NGC~1068 prompted  
Pier \& Krolik (1993) to suggest that there was an additional clumpy 
dust component above the torus.  This two component
torus+cone model
was also used by Efstathiou et al. (1995)  to match the broad
spectrum of NGC~1068.  

To suppress the silicate emission in type 1 objects, previous models
forced the torus geometry to have extremely steep inner regions so
that the region directly illuminated by UV radiation could
only be viewed at high inclination angles.
Since the models were also driven by the belief that 
there was strong silicate absorption in the Seyfert 2s, the outer disks
were arranged so that at high inclination angles the inner
hotter regions were only viewed through colder material
(e.g., Pier \& Krolik 1993; Efstathiou \&
Rowan-Robinson 1995; Granato et al. 1997).   Also
Pier \& Krolik (1993) models assumed high optical depths to 
be consistent with the high hard X-ray column densities often
measured in Seyfert 2 galaxies.

Clavel et al. (2000) have used {\it ISO} spectroscopy 
(through a $24\arcsec \times 24\arcsec$ aperture) to
show that Seyfert 1 galaxies 
have little silicate emission, and a broad
smooth infrared continuum.  However, the Seyfert 2s displayed PAH emission
and lower levels of continuum.   In fact recent {\it ISO} and 
ground-based mid-infrared spectroscopy of a small 
number of nearby AGN show that the silicate 
absorption feature in some Seyfert 2s might be shallower than previously 
inferred (see e.g., Laurent et al. 2000, Imanishi \&
Ueno 2000 and Le Floc'h et al. 2001) because measurements of the nearby continuum 
were contaminated with PAH features originating in the host galaxy. 
The mid-infrared emission is probably slightly  extended in many Seyfert 2s 
(Maiolino et al. 1995) across the region covered 
by the {\it ISO} spectra, 
so it is likely that there is some PAH contribution from the 
surrounding galaxy in a number of cases.
The nuclear spectrum itself might contain little or no PAH emission.
For example, little silicate absorption is seen in the {\it ISO}/SWS nuclear spectrum  
of NGC~1068 and the continuum is nearly featureless (Le Floc'h al. 2000, 
see also our Figure~2, left panel). 

In addition to the behavior of silicate feature, torus models
can be constrained by the shapes of the SEDs for a complete sample of galaxies. 
To do so, we use the tapered 
disk model of Efstathiou \& Rowan-Robinson (1995). 
In this geometry, the height of the disk
increases with the radial distance, but tapers off to a constant
height in the outer part. The main parameters of the model
are the half opening angle of the torus $\Theta_{\rm c}$, 
the equatorial UV optical depth $\tau_{\rm UV}$, the dust 
sublimation temperature $T$, the ratio
between inner and outer radii $r_1/r_2$, and the 
ratio between the scale height and the outer radius $h/r_2$. 
Table~8 summarizes the parameters of the models used here. 
In addition to the torus models, we will also consider the composite
torus+cone model used by Efstathiou et al. (1995) 
to fit the SED of NGC~1068. 
The viewing angles are defined as $\theta_{\rm v} =0\arcdeg$ for 
the polar view of the AGN, and 
$\theta_{\rm v} = 90\arcdeg$ for the equatorial view of the AGN 
(note that in the Efstathiou \& Rowan-Robinson
1995 models the authors measured the angles from the equator).

Because of their high equatorial opacities, 
the torus models presented here (but also those 
of Pier \& Krolik 1993;
Granato \& Danese 1994; Granato et al. 1997) generally
predict the following: 
(i) for viewing angles $\theta_{\rm v} < \Theta_{\rm c}$
an SED more or less flat in $\nu f_\nu$ depending on the 
model input parameters, and a direct view of the nuclear optical
emission; (ii) for viewing angles 
$\theta_{\rm v} > \Theta_{\rm c}$, a steep near 
and mid-infrared SED becoming
steeper as the viewing angle increases, with no direct view of the nucleus; 
(iii) an intermediate SED as well as some nuclear 
optical emission for only a very small range of possible viewing angles
around $\theta_{\rm v} \simeq \Theta_{\rm c}$.
A dichotomy of spectral shapes
is thus predicted, and few, if any, Seyferts are expected to have 
an intermediate SED. This behavior is shown graphically in Figure~9 where 
the predicted infrared spectral index ($\alpha_{\rm IR}$) is plotted against 
the cosine of the viewing angle for some of the models described 
in Table~8.

\section{Comparison with observations}
The observed 
distribution of $\alpha_{\rm IR}$ (see Figure~6) in the 
extended CfA sample clearly lacks the dichotomy of
spectral indices predicted 
by the models. As discussed in Section~5, all the Seyfert $1-1.5$s
and a large fraction of the Seyfert $1.8-1.9$s show infrared spectral indices 
clustered near $\alpha_{\rm IR}=1.5\pm0.3$. Type 
2 objects on the other hand show a 
variety of steeper spectral indices, unlike the model predictions. 
Moreover the large opacity torus models without the extended cone component 
predict steeper SEDs ($\alpha_{\rm IR} > 4$) than observed 
in the CfA Seyferts for sufficiently large viewing angles. Below, 
we discuss a number of possibilities to reconcile the observed SEDs with 
the model predictions.

\subsection{Torus models with low opacities}

Figure~9 shows  
that even the model with an equatorial optical depth of 
$\tau_{\rm UV}=250\,$mag ($A_V=50\,$mag) produces SEDs with 
$\alpha_{\rm IR} > 4$ 
for viewing angles $\theta_{\rm v} > 60\arcdeg$. The other torus models 
considered here with $\tau_{\rm UV} > 250\,$mag 
predict even steeper SEDs near an equatorial view of the AGN. 
Granato et al. (1997) and Fadda et al. (1998) used similar arguments
based on the observed near and mid-infrared colors of Seyfert 
galaxies to favor low opacity (equatorial 
optical depths $A_V \simeq 10-30\,$mag) for extended 
(a few hundred parsec) torus models.

Traditionally low opacity torus models have been 
ruled out because they tend to produce 
the $9.7\,\micron$ silicate feature in  
emission for viewing angles of less than the half opening angle 
of the torus (that is, type 1 objects).  Ground-based
mid-infrared spectroscopy of 
Seyfert 1s did not show evidence for this feature 
in emission (Roche et al. 1991). Recently, however, 
Clavel et al. (2000) have argued  
that the average {\it ISO} mid-infrared spectrum of 
a sample of Seyfert 1 galaxies  displays some evidence for 
the presence of the silicate feature in emission. Spoon et al. (2002)
on the other hand do not find evidence for this feature. 
Additional observations are needed to resolve this issue. 

\subsection{Torus models with an extended (optically thin) component}
 
The extended conical component proposed in Efstathiou et 
al. (1995) was assumed to be optically 
thin and thus increases the continuum level around $3-5\,\micron$, 
causing the infrared spectral index to be shallower even 
for high viewing angles (see Figure~9). 
Indeed, NGC~1068 shows emission by dust 
from the wall of the cavity, i.e., perpendicular to the axis of the torus 
and in an extended component
(100\,pc) coincident with the radio emission and ionization cone 
(see Bock et al. 1998). 
The Torus+Cone model of Efstathiou et al. (1995) seems to 
reproduce well the observed distribution of $\alpha_{\rm IR}$ in the
type 2 objects in the CfA sample, 
although it will be necessary to increase the opening 
angle of the torus to match the observed relative numbers of type 1 and 
type 2 objects in the CfA sample. This model
also reproduces well the observed infrared spectral indices 
of Seyfert $1-1.5$s. 

This model still fails to solve  the problem of  
the intermediate Seyferts, as infrared spectral 
indices in the range $\alpha_{\rm IR} = 1.5\pm0.1-2.5\pm0.1$ 
are extremely unlikely to occur. This discrepancy can be solved with fine tuning 
of the models. For instance, a torus+cone model 
with a lower equatorial optical depth would make 
the transition of infrared spectral indices 
between type 1 and type 2 smoother, thus narrowing the 
range of 'unlikely' spectral indices.  Another 
way of producing intermediate SEDs is having 
a range in the visibility of the
scattering region. This could be done with a range  in the
$h/r_2$ parameter or if  the scattering region is  physically located at
different distances from the nucleus. A more visible scattering region
would produce  a  stronger  $1-2\,\micron$  continuum and therefore 
shallower SEDs. This could  account for some of the intermediate Seyferts
(Figure~4).

\subsection{Clumpy torus models}
Nenkova, Ivezi\'c, \& Elitzur (2002) have recently put forward 
a more realistic model for the emission by dust in AGNs following 
the initial suggestion by Pier \& Krolik (1993) that the dust in the torus 
must be in clumps to protect the grains.  The parameters of this 
clumpy torus are, 
apart from the usual geometry parameters of the torus, the radial 
distribution of clouds, the number of clouds and the optical depth 
of the individual clouds. In their model a cloud is heated by radiation 
from both the AGN and other surrounding clouds. 
The main result from this 
paper is that this model can reproduce simultaneously the broadness of
the infrared SEDs and suppress the silicate feature emission with a 
few clouds ($5-10$) with 
optical depths $\tau_V > 60$.

Nenkova et al. (2002) only showed results from the limiting 
viewing angles of $\theta_{\rm V} = 0\arcdeg$ and 
$\theta_{\rm V} = 90\arcdeg$, but from their figure~3
we can see that the equatorial view of the AGN in that particular
model produces an SED consistent 
with the steepest SEDs in the CfA sample.  
It remains to be seen if their model could 
reproduce the variety of observed SEDs measured for type 2 objects
in the CfA sample.  One concern in their comparison with observations, although 
they will provide a more detailed modelling in the near future, is the small
number of type-2 SEDs fitted, and the possibility that some of the examples 
are contaminated by stellar processes. For example, for Circinus, 
one of the Seyfert 2 galaxies used in their modelling, 
Ruiz et al. (2001) demonstrate that a dominant starburst component is present 
at wavelengths longer than approximately $25\,\micron$. 

\subsection{Dust composition}

The lack of deep silicate absorption 
features in Seyfert 2s and other arguments based on 
anomalous $A_{\rm V}/N_{\rm H}$ ratios in AGNs have prompted the 
idea that the composition and properties of dust in AGNs may be 
different from those of Galactic dust (see e.g., Laor
\& Draine 1993; Maiolino et al. 2001; Maiolino, Marconi,
\& Oliva 2001). 

Specifically, the dust around AGNs is proposed to have a large portion of large grains
(Maiolino et al. 2001; Maiolino, Marconi, \& Oliva 2001). 
In that case, the extinction curve would
be flatter than the standard Galactic extinction curve.
A model with a higher fraction of large grains (or fewer silicates) means 
we can have a flat type 1 spectrum (with
weak or no silicate emission feature) with a lower optical depth, as well
as shallower SEDs for viewing angles $\theta_{\rm v}
> \Theta_{\rm c}$ (that is, type 2s). Fadda et al. (1998) have shown
the effects on near infrared and mid infrared colors using
models with fewer silicate grains than the standard Galactic dust 
composition. Such models would be in better agreement with our findings for 
the extended CfA sample.

\section{Summary}

We  define an extended CfA sample of Seyfert galaxies 
composed of the original CfA Seyferts and those 
galaxies initially classified as LINERs, but later reclassified
as Seyferts from higher S/N optical 
spectra. For this sample, we present high resolution 
IRTF $K^\prime$ ($2.2\,\micron$) and $L$-band ($3.5\,\micron$)
images and compile {\it HST}, {\it ISO} 
and ground-based observations in the 
spectral region $0.4-16\,\micron$. These data have been used to determine 
the unresolved nuclear emission with improved
accuracy over any previous studies. 

The resulting spectral energy distributions are used to
test and constrain unified models for the dusty tori in the active nuclei.
We find that the fitted infrared spectral indices 
($f_\nu \propto \nu^{-\alpha_{\rm IR}}$) between
1 and $16\,\micron$ are correlated with the Seyfert type in the sense that
steeper SEDs tend to be found in Seyfert 2s and flatter ones
in Seyfert $1-1.5$s.  The majority
of galaxies optically classified as Seyfert types 1.8 and 
1.9 display infrared spectral indices and SEDs either 
similar to those of type 1 objects or
intermediate between Seyfert 1s and Seyfert 2s. The 
intermediate SEDs of many Seyfert $1.8-1.9$s may be consistent with 
the presence of a pure Seyfert 1 viewed through a moderate amount 
($A_V \lesssim 5\,$mag) of foreground galaxy extinction.

However, the abrupt division in SED shape correlated with emission-line
properties that would be predicted by an optically
thick torus model ---flat for type 1s and steep for type 2s--- 
is not observed. Discrepancies with this model include the wide 
range of spectral indices observed in type 2 objects, 
the lack of extremely steep SEDs, and the large number of objects with intermediate properties. 
In addition, there are a surprisingly large number of galaxies with
broad-line components and steep SEDs, a combination that conventional
torus models would predict is very rare. 

These results, combined with the lack of deep $9.7\,\micron$ silicate absorption
feature in Seyfert 2s and the possible presence of silicate
emission in Seyfert 1s (Clavel et al. 2000), suggest
that the region of infrared emission might not be as optically thick
as previously considered. 
Although X-ray and maser observations show 
that high opacity material does exist near the AGN,
it is not clear whether this 
material is responsible for the infrared emission.
In a recent paper, Risaliti, Elvis, \& Nicastro (2002) have shown
that the observed fast variability (a few months) of hard X-ray 
column densities in Seyfert 2s is not consistent with the absorbing material being located in 
the standard parsec scale torus. Since this material may be closer to the black hole
than the region responsible for the infrared emission in Seyfert galaxies, the hard X-ray column 
densities might not be useful for constraining the torus optical depth. 

Other possible explanations for the observed SEDs
include an optically thick torus with 
an additional optically thin infrared component (such as from
a cone of dust clouds: Efstathiou et al. 1995) to  account
for the lack of very steep SEDs and the lack of deep silicate 
absorption features.  Another promising approach is the clumpy 
torus model of Nenkova et al. (2002).

\acknowledgments

We are grateful to Emeric Le Floc'h for providing us with 
the {\it ISO} nuclear spectrum of NGC~1068, and to Paul Martini, 
Maia Nenkova and 
Moshe Elitzur for stimulating conversations. 
We also thank the IRTF staff for their support during the observing
campaigns.
Support for this work was provided by NASA through grant number
GO-07869.01-96A
from the Space Telescope Institute, which is operated by the Association
of Universities for Research in Astronomy, Incorporated, under NASA
contract NAS5-26555.
We also acknowledge support from NASA projects NAG-53359 and
NAG-53042 and from JPL Contract No.~961633.

This research has made use of the NASA/IPAC Extragalactic Database (NED)
which is operated by the Jet Propulsion
Laboratory, California Institute of Technology, under contract with
the National Aeronautics and Space Administration.


\pagebreak

\section*{Figure Captions}

{\bf Figure~1.---} Contours plots of selected galaxies with 
extended emission in the $L$-band (right panels). We 
show the  $K^\prime$-band contour plots for comparison (left panels).
The orientation of the images is south up, east to the left.

{\bf Figure~2.---} {\it Left panel:} Non-stellar SED of NGC~1068 in units of 
erg\,cm$^{-2}$\,s$^{-1}$. The {\it ISO} spectrum 
of the AGN (nuclear) component of NGC~1068 
is from Le Floc'h et al. (2001).
The additional mid-infrared data points
at 12.6, 17, 22.5, 24.5, and $33.5\,\mu$m
are ground-based small aperture observations from Rieke \& Low (1975a); 
data points at 11.2 and $20.5\,\micron$ are from Alloin et 
al. (2000). {\it Right panel:} Non-stellar SED of NGC~4151 in units of 
erg\,cm$^{-2}$\,s$^{-1}$. The additional mid-infrared data points
at 10 and $18\,\mu$m (unresolved emission) are from Radomski et 
al. (2002), at $21\,\mu$m is from Rieke \& Low (1972) and 
at $33.5\,\mu$m from Rieke \& Low (1975b).

{\bf Figure~3.---} Observed non-stellar 
SEDs ($\nu\,f_{\nu}$)
of optically classified  type 1 ($1-1.5$) CfA Seyferts  in 
units of erg\,cm$^{-2}$\,s$^{-1}$.

{\bf Figure~4.---} Observed non-stellar 
SEDs ($\nu\,f_{\nu}$) of 
optically classified intermediate type ($1.8-1.9$) CfA Seyferts in 
units of erg\,cm$^{-2}$\,s$^{-1}$.

{\bf Figure~5.---} Observed non-stellar 
SEDs ($\nu\,f_{\nu}$) of optically classified type 2 CfA Seyferts in 
units of erg\,cm$^{-2}$\,s$^{-1}$.

{\bf Figure~6.---} Distributions of the
fitted spectral indices ($f_\nu \propto \nu^{-\alpha}$) 
in the optical-infrared wavelengths 
($\alpha_{\rm opt-IR}$ for the 
spectral range $\simeq 0.4-16\,\micron$,  left panels) 
and infrared wavelengths ($\alpha_{\rm IR}$ for the 
spectral range $1-16\,\micron$, right panels). 
The very top panels are the 51 Seyferts in the 'extended' 
CfA sample with measured spectral indices. The panels below are the same sample
broken up in different Seyfert types as derived 
from optical spectroscopy.

{\bf Figure~7.---} Effects of foreground extinction ($A_{\rm V} = 0.4, \, 1\,
{\rm and} \, 5\,$mag) on the observed SED of 
an AGN, using Rieke \& Lebofsky (1985) extinction 
law. The AGN SEDs are outputs of the Torus+Cone model 
from Efstathiou \& Rowan-Robinson 1995 
(see Section~6 and Table~8 for details) for three 
different viewing angles. 
$\theta_{\rm v} = 0\arcdeg$ is the polar view of the AGN, 
$\theta_{\rm v} = 30\arcdeg$ is an intermediate view (corresponding
to the half opening angle of the torus, $\Theta_{\rm c}$), and 
$\theta_{\rm v} = 90\arcdeg$ is the 
equatorial view (through the torus).

{\bf Figure~8.---} Fitted optical-infrared (left panel) and 
infrared (right panel) spectral indices (see Section~5) 
as a function of the inclination
of the host galaxy, where $b/a=0$ is an edge-on galaxy, and 
$b/a=1$ is a face-on galaxy. The arrows at the top of the plots
indicate the values of $b/a$ for the remaining 7 galaxies with no measured 
spectral indices ---dotted lines for the type 2s (green 
in the on-line version) and solid lines for the type 
1s (red in the on-line version). The dashed line indicates a host galaxy 
axial ratio of $b/a=0.6$ or an inclination
of $i> 53\arcdeg$ (see discussion in text). The galaxies 
are separated between Seyfert 1s ($1-1.5$) shown as open circles (red in
the on-line version) and 
Seyfert 2s ($1.8-2$) shown 
as filled star symbols (green in the on-line version), 
using the  classification from optical spectroscopy.

{\bf Figure~9.---} Predicted infrared spectral 
indices ($f_\nu \propto \nu^{-\alpha_{\rm IR}}$) 
in the $1-16\,\micron$ spectral 
range as a function of the cosine of the viewing angle to the torus 
($\theta_{\rm v} = 0\arcdeg$
polar view of the AGN, $\theta_{\rm v} = 90\arcdeg$
equatorial view of the AGN)  using 
Efstathiou \& Rowan-Robinson (1995) and 
Efstathiou et al. (1995) torus models. {\it Left panel:} 
results for models with a torus half opening angle
$\Theta_c =45\arcdeg$, and two different equatorial UV 
optical depths: $\tau_{\rm UV} = 500\,$mag (filled triangles),  
$\tau_{\rm UV} = 250\,$mag (filled stars), Models Torus 2 and Torus 3 
respectively (see Table~8). {\it Right panel:} results for 
a torus model with a half opening angle $\Theta_c =30\arcdeg$, and  
an equatorial UV optical depth of $\tau_{\rm UV} = 1200\,$mag
with the cone component (Model Torus+Cone in Table~8, open circles) and 
without the cone component
(Model Torus 4 in 
Table~8, filled circles). In both diagrams the shaded areas
represent the observed ranges of fitted $\alpha_{\rm IR}$ for
Seyfert $1-1.5$s (darkest shaded 
area), Seyferts $1.8-1.9$s (intermediate shaded area), 
and Seyfert 2s (lightest shaded area) in the CfA sample.

\pagebreak


\begin{deluxetable}{ccccccccc}

\tiny
\tablewidth{18cm}
\tablecaption{Summary of high resolution optical, near-infrared and 
mid-infrared observations for the CfA Seyfert galaxies.}
\tablehead{
\multicolumn{1}{l}{Galaxy}      &
\colhead{Type}                 &
\colhead{WFPC1/2}                  &
\colhead{Ref}            & 
\colhead{NICMOS}          &
\colhead{Ref}         &
\colhead{$KL$}        &
\colhead{Mid-IR} & \colhead{Ref}\\
\colhead{(1)}                        & 
\colhead{(2)}                        & 
\colhead{(3)}                        & 
\colhead{(4)}                        & 
\colhead{(5)}                        &
\colhead{(6)}                        &
\colhead{(7)}                        &
\colhead{(8)}                        &
\colhead{(9)}} 
\startdata
Mkn~334 & Sy1.8 &* & 1 & F110W, F160W & 2, 4 & 2 & 6.75, 9.63, 10, 16$\micron$ 
& 5, 9, 6\\
Mkn~335 & Sy1.0 & F785LP & 1 &\nodata & \nodata & 2 & 6.75, 9.63, 16$\micron$ 
& 5, 6\\
UGC~524/0048+29 & Sy1.0 & \nodata       & \nodata  &\nodata & \nodata & \nodata 
& 10$\micron$ & 9\\
I~Zw~1  & Sy1.0 & F555W      & 1  & \nodata &\nodata& 10      & 
5.9, 7.7, 10, 16$\micron$ & 7, 10, 6\\
Mkn~993 & Sy$1.5-2.0$ & F606W & 1 & \nodata & \nodata& 2 & 10, 16$\micron$ & 11, 6\\
\\
Mkn~573 & Sy2.0 (non-HBLR) & F708W & 2 & F110W, F160W & 2, 4& 3 & 10, 16$\micron$ & 11, 6\\
UM~146/UGC~1395 & Sy1.9 & F606W & 1  &F110W, F160W  & 2, 4 & 2 & 6.75, 10, 16$\micron$ & 
2, 9, 6\\
Mkn~590/NGC~863 & Sy1.0 & F785LP & 1  & \nodata & \nodata & 2 & 6.75, 9.63, 16$\micron$ 
& 5, 6\\
NGC~1068 & Sy2.0/Sy1.8 (HBLR) & \nodata & \nodata & F110W, F160W & 3 & 3 &
$5-16$, 16$\micron$ & 13,6\\
         &                   &         &           & F222M\\
NGC~1144& Sy2.0 (non-HBLR) & F606W & 2 & F110W, F160W & 2, 4& 2 & 6.75, 10, 16$\micron$ & 2, 11, 6\\
\\
Mkn~1243/NGC~3080 & Sy1.0 & \nodata & \nodata & \nodata  &\nodata & \nodata   & 10, 16$\micron$ & 9, 6\\
NGC~3227 & Sy1.5 & F547M & 1 & F160W, F222M & 3 & 2 & 6.75, 9.63, 16$\micron$ 
& 5, 6\\
NGC~3362 & Sy2.0 (non-HBLR) & F606W & 2 &F110W, F160W &2, 4 & 2 & 10, 
16$\micron$ & 9, 6\\
UGC~6100 & Sy2.0 (non-HBLR) & F606W & 2 &F110W, F160W & 2, 4 & 2 & 6.75, 9.63, 10, 16$\micron$ 
& 5, 9, 6\\
NGC~3516 & Sy1.5   & F547M & 1 & F160W &4 & 14 & 6.75, 9.63, 16$\micron$ 
& 5, 6\\
\\
NGC~3786/Mkn~744 & Sy1.8 &* & 1 & F110W, F160W & 2, 4 &2 & 
10, 16$\micron$ & 9, 6\\
NGC~3982 & Sy2.0/Sy1.9 (non-HBLR)& F606W & 2 &F110W, F160W  & 2, 4 & 2 & 6.75, 9.63, 10, 16$\micron$ 
& 5, 9, 6\\
NGC~4051 & Sy1.0 & F547M & 1 & \nodata &\nodata & 2 & 6.75, 9.63, 10, 16$\micron$ 
& 5, 10, 6\\
NGC~4151 & Sy1.5   & F547M & 1 & F110W, F160W& 3 & 3 & 5.9, 7.7, 10, 16$\micron$ & 7, 10, 6\\
         &         &       &   & F222M\\
NGC~4235 & Sy1.0 & F547M & 1 & F160W &4 & 2 & 10, 16$\micron$ & 9, 6\\
\\
NGC~4253/Mkn~766 & Sy1.5 & F785LP & 1 & F160W & 4& 2 & 6.75, 9.63, 10,  
16$\micron$ & 5, 9, 6\\
Mkn~205 & Sy1.0 & F814W  & 1  &\nodata   & \nodata & 10  & 6.75, 10, 16$\micron$ & 2, 16, 6\\
NGC~4388 & Sy2.0/Sy1.9 (HBLR) & \nodata & \nodata & F110W, F160W & 2, 4 &8 & 6.75, 9.63, 10, 16$\micron$ 
& 5, 17, 6\\
NGC~4395 & Sy1.0/Sy1.8 & F450W & 1 & F160W & 4 & 2 & 10, 12$\micron$ & 9, 15 \\
Mkn~231 & Sy1.0 & F814W & 1 & F110W, F160W & 2, 4 & 10 & 5.9, 7.7, 10, 16$\micron$ & 7, 10, 
6\\
\\
NGC~5033 & Sy1.9/Sy1.5 & F547M & 2 & F110W, F160W & 2, 4 & 2 & 6.75, 9.63, 10, 16$\micron$ & 5, 9, 6\\
UGC~8621/1335+39 & Sy1.8   & \nodata  & \nodata & \nodata & \nodata  & \nodata & 
6.75, 9.63, 16$\micron$ & 5, 6\\
NGC~5252 & Sy1.9 (HBLR) & F547M & 2 & F110W, F160W & 2, 4 &3 & 6.75, 9.63, 10, 16$\micron$ & 3, 9\\
NGC~5256/Mkn~266SW & Sy2.0 (non-HBLR) & \nodata & \nodata &F110W, F160W  & 
2, 4 & 2 & 
6.75, 9.63, 10, 16$\micron$  & 5, 9, 6\\
NGC~5283/Mkn~270 & Sy2.0 (non-HBLR)& F708W & 2 &F110W, F160W  & 2, 4 & 2 & 
6.75, 10, 16$\micron$ & 2, 10, 6\\
\\
NGC~5273 & Sy1.9/Sy1.5 & F439W & 2 &F110W, F160W  & 2, 4 & 2 & 6.75, 9.63, 10, 
16$\micron$ & 5, 9, 6\\
Mkn~461  & Sy2.0 & \nodata  & \nodata & F110W, F160W & 2, 4 & 2& 6.75, 10$\micron$ & 2, 9 \\
NGC~5347 & Sy2.0 (non-HBLR) & \nodata & \nodata &F110W, F160W  & 2, 4 & 2 &
$10\micron$ & 9 \\
Mkn~279 & Sy1.0   & * & 1& \nodata & \nodata & 10 & 6.75, 9.63, 16$\micron$ 
& 5, 6 \\
NGC~5548 & Sy1.5 & F785LP & 1 & F160W, F222M & 3 & 2 & 6.75, 
9.63, 16$\micron$ & 5, 6\\
\\
Mkn~471  & Sy1.8 & F606W & 1 & F110W, F160W & 2, 4 & 10 & $10\micron$ & 10  \\
NGC~5674 & Sy1.9 & F606W & 1 &F110W, F160W  & 2, 4 & 2 & 6.75, 9.63, 10$\micron$ 
& 5, 9\\
Mkn~817 & Sy1.5 & F785LP & 1 & \nodata & \nodata & 12 & 6.75, 9.63, 16$\micron$ 
& 5, 6\\
NGC~5695/Mkn~686 & Sy2.0 (non-HBLR) &F606W & 2 & F110W, F160W & 2, 4 & \nodata & 10, 12$\micron$ 
& 9, 2\\
Mkn~841  & Sy1.5 & F785LP & 1 & \nodata & \nodata  & 2 & 6.75, 9.63, 10, 16$\micron$ 
& 5, 9, 6\\
\tablebreak
\\
NGC~5929 & Sy2.0 (non-HBLR)& F606W & 2 &F110W, F160W  & 2, 4 & 2 & 6.75, 9.63, 10, 16$\micron$ & 5, 9, 6\\
NGC~5940 & Sy1.0 & F606W & 1 & F160W & 4 & 2 & 6.75, 9.63, 10$\micron$ 
& 2, 9\\
Mkn~1513 & Sy1.0 &\nodata & \nodata &\nodata & \nodata & 2  & 6.75, 10, 14.3$\micron$ & 2, 10\\
NGC~6104 & Sy$1.5-1.8$   & F606W & 1 & F160W & 4 & \nodata  & 10, 16$\micron$ & 9, 6\\
UGC~12138 & Sy1.8 & * & 1 &F110W, F160W  & 2, 4 & 2 & 6.75, 9.63, 10, 16$\micron$ & 2, 9, 6 \\
\\
NGC~7469 & Sy1.0 & F785LP & 1 &F110W, F160W & 3 & 2&
6.75, 10, 16$\micron$ & 8, 10, 6 \\
         &       &        &   & F222M\\
NGC~7603/Mkn~530 &  Sy1.5/Sy1.8 & F785LP & 1 & \nodata & \nodata & 2 & 6.75, 
9.63, 10$\micron$ & 5, 10\\
NGC~7674/Mkn~533 & Sy2.0 (HBLR) & \nodata & \nodata & F110W, F160W & 
2, 4 & 3 & 6.75, 9.63, 10,  
16$\micron$ & 5, 9, 6\\
NGC~7682 & Sy2.0 (HBLR) & F606W & 2 &F110W, F160W  & 2, 4 & 2 & 6.75, 10, 
16$\micron$ & 2, 9, 6\\

\enddata
\tablecomments{Column~(2): Seyfert types  taken from Osterbrock 
\& Martel (1993) and Ho et al. (1997). NGC~6104 has 
been classified as a Seyfert $1.5-1.8$ galaxy by Jim\'enez-Benito
et al. (2000). Goodrich (1995) reports that NGC~7603/Mkn~530 and Mkn~993 
(see also Tran, Osterbrock, \& Martel 1992) have 
undergone transitions between types Seyfert 1 and Seyfert $1.8-1.9$.
We also indicate in this column 
for the Seyfert 2s whether hidden broad line regions have been detected from 
spectropolarimetry (HBLR) or not (non-HBLR). Column~(3): {\it HST} optical 
filter, unless the *
symbol is used in which case the optical unresolved flux
was estimated from ground-based
observation (see Ho \& Peng (2001) for more details). Column~(4):  
references for the unresolved optical fluxes. Column~(5): {\it HST}/NICMOS filters.
Column~(6): references for 
the NICMOS fluxes. Column~(7): references for the $KL$-band fluxes.
Column~(8): Mid-infrared data available. 5.9, 6.75, 7.7, 9.63, 12.5, 14.3 and $16\,\mu$m
indicate {\it ISO} data, and $10\,\mu$m indicates ground-based $N$-band measurements. 
Column~(9): references for the mid-infrared fluxes.
\\
References in Columns~(4), (6), (7) and (9): 1. Ho \& Peng (2001). 2. This work.
3. Alonso-Herrero et al. (2001). 4. Quillen et al. (2001) 5. Clavel et al. 
(2000). 6. P\'erez Garc\'{\i}a \& Rodr\'{\i}guez Espinosa (2001). 
7. Rigopoulou et al. (1999). 8. Alonso-Herrero et al. (1998). 9. Maiolino et al.
(1995). 10. Rieke (1978). 11. Edelson et al. (1987). 
12. Rudy, LeVan, \& Rodr\'{\i}guez-Espinosa (1982). 13. Le Floc'h, 
private communication (2002). 14. McAlary, McLaren, \& Crabtree (1979).
15. Bendo et al. (2002). 16. Neugebauer et al. (1979). 17. Scoville et al.
(1983).}
\end{deluxetable}

\begin{deluxetable}{ccccccccc}
\scriptsize
\tablecaption{Summary of high resolution optical, near-infrared and 
mid-infrared observations for additional Seyfert galaxies in 
the CfA sample.}
\tablehead{
\multicolumn{1}{l}{Galaxy}      &
\colhead{Type}                 &
\colhead{WFPC1/2}                  &
\colhead{Ref}            & 
\colhead{NICMOS}          &
\colhead{Ref}            &
\colhead{$KL$ band}        &
\colhead{Mid-infrared} & \colhead{Ref}\\
\colhead{(1)}                        & 
\colhead{(2)}                        & 
\colhead{(3)}                        & 
\colhead{(4)}                        & 
\colhead{(5)}                        &
\colhead{(6)}                        &
\colhead{(7)}                        &
\colhead{(8)}                        &
\colhead{(9)}} 
\startdata
NGC~3031 & Sy1.5 & F547M & 1 & F160W & 2 & 9 & 10\,$\micron$ & 10\\
NGC~3079 & Sy2.0 & \nodata & \nodata & F160W & 2 &9
& 10, 16\,$\micron$ & 6, 7 \\
NGC~3185 & Sy2.0: & \nodata & \nodata & \nodata & \nodata &
\nodata & 10, 12\,$\micron$ & 11, 4\\
NGC~4258 & Sy1.9 & F547M & 1 & F110W, F160W, F222M & 3 & 
3 & 10.5, 12.5, 17.9\,$\micron$ & 3\\
NGC~4501 & Sy2.0 & \nodata & \nodata & \nodata & \nodata &9 & 
10\,$\micron$ & 6\\
NGC~4579 & Sy1.9/L1.9 & F547M & 1 & \nodata & \nodata  &4 & 6.75, 9.63$\micron$ 
& 5 \\
NGC~5194 & Sy2.0 & \nodata & \nodata & F110W, F160W, F222M & 2, 4 & 8
& 9.55, 12.5\,$\micron$ & 4 \\
NGC~7479 & Sy1.9 & F569W & 1 & F160W & 2 &4 & 
6.75, 10, 14.3\,$\micron$ & 4, 9 \\
NGC~7743 & Sy2.0 & \nodata & \nodata & F160W & 2 & 4 & \nodata&
\nodata\\
\enddata
\tablecomments{See Table~1 for an explanation. Column~(2): the Seyfert 
types are from  Ho et al. (1997). Column~(8): the mid-IR fluxes 
of NGC~4258 are ground-based.\\
References in Columns~(4), (6), (7) and (9): 1. Ho \& Peng (2001). 
2. Quillen et al. (2001) 3. Chary 
et al. (2000). 4. This work. 5. Clavel et al. 
(2000). 6. Devereux (1987). 
7. P\'erez Garc\'{\i}a \& Rodr\'{\i}guez Espinosa (2001). 
8. Ellis,  Gondhalekar, \& Efstathiou (1982). 9. Willner et al. (1985).
10. Rieke \& Lebofsky (1978). 11. Maiolino et al. (1995).} 
\end{deluxetable}

\begin{deluxetable}{cccc}
\tablewidth{9cm}
\tablecaption{Distribution of type 1s and type 2s in different
samples.}
\tablehead{
\colhead{Sample}         &
\colhead{No. galaxies}   &
\colhead{Type 1}      &
\colhead{Type 2}      \\
\colhead{(1)}                        & 
\colhead{(2)}                        & 
\colhead{(3)}                        & 
\colhead{(4)}} 

\startdata
\multicolumn{4}{c}{All $b/a$}\\
\hline
Extended CfA  & 58 & 43\% & 57\% \\
RSA           & 60 & 28\% & 72\% \\
Palomar$^*$   & 44 & 27\% & 73\% \\
$12\,\micron$ & 55 & 45\% & 55\% \\
\hline
\multicolumn{4}{c}{$b/a \ge 0.6$ }\\
\hline
Extended CfA  & 47 & 43\% & 57\%\\
RSA           & 41 & 37\% & 63\%\\
Palomar$^*$   & 32 & 28\% & 72\%\\
$12\,\micron$ & 39 & 56\% & 44\%\\
\enddata
\tablecomments{Distribution of type 1 ($1-1.5$) and type 2 
($1.8-2$) Seyferts
based on their optical classification.\\
$^*$ Only galaxies with certain Seyfert classification are included.}
\end{deluxetable}

\begin{deluxetable}{ccccccc}
\footnotesize

\tablewidth{15cm}
\tablecaption{IRTF $K^\prime L$-band Aperture Photometry 
and L-band unresolved contribution 
within a 3\arcsec-diameter aperture.}
\tablehead{
\multicolumn{1}{l}{Galaxy}      &
\colhead{FWHM}                 &
\colhead{Aperture}                  &
\colhead{$K^\prime$}          &
\colhead{$K^\prime-L$}      &
\colhead{$\frac{f_{\rm unresolved}}{f_{3''}}$}  &
\colhead{$L$-band Morphology}
\\
\colhead{(1)}                        & 
\colhead{(2)}                        & 
\colhead{(3)}                        & 
\colhead{(4)}                        & 
\colhead{(5)}                        &
\colhead{(6)}                        &
\colhead{(7)}} 
\startdata
Mkn~334         &  0.78 & 1.5 & 11.64 & 1.53 & 89\% & Compact\\
                &       & 3   & 11.21 & 1.44 \\
                &       & 6   & 10.97 & 1.32 \\
	        &       & 9   & 10.83 & \nodata\\
Mkn~335         &  0.72 & 1.5 & 10.88 & 1.76 & 94\% & Compact\\  
                &       & 3   & 10.60 & 1.76 \\ 
                &       & 6   & 10.50 & 1.72 \\
	        &       & 9   & 10.46 & \nodata\\
UM~146/UGC~1395 &  0.84 & 1.5 & 13.19 & 1.21 & 65\% & Compact \\
                &       & 3   & 12.63 & 1.10 \\
                &       & 6   & 12.25 & 0.82 \\
                &       & 9   & 12.00 & \nodata\\
Mkn~590         &  0.86 & 1.5 & 11.96 &  2.25 & 76\% & Compact \\
                &       & 3   & 11.38 & 1.95 \\ 
                &       & 6   & 11.00 & 1.68 \\
                &       &     & 10.80 & \nodata\\
Mkn~993         &\nodata & 6  & 11.40 & 0.36 & \nodata & Blind offsets\\
NGC~1144        &  1.11 & 1.5 & 13.19 & 1.14 & 40\% & Extended\\ 
                &       & 3   & 12.22 & 0.84\\
                &       & 6   & 11.51 & 0.60\\
                &       & 9   & 11.05 & 0.52\\
UGC~6100        &\nodata & 6   & 12.23 & 0.23 &  \nodata & Blind offsets\\ 
NGC~3227        & 0.81  & 1.5 & 10.87 & 1.37 &  71\% & Compact\\
                &       & 3   & 10.24 & 1.15 \\
                &       & 6   & 9.82  & 0.91 \\
                &       & 9   & 9.62  & 0.76\\
NGC~3362        & \nodata   & 6   & 12.80 & 0.40 & \nodata & Blind offsets\\
NGC~3786/Mkn~744 & 1.08 & 6   & 10.74 & 1.07 & 56\% & Tracking problems \\
NGC~3982        & \nodata & 6   & 12.67 & 1.17 & \nodata & Blind Offsets\\
NGC~4051        & 0.78  & 1.5 & 10.54 & 1.59 & 73\% & Compact\\
                &       & 3   & 10.08 & 1.44 \\
                &       & 6   & 9.80  & 1.30 \\
                &       & 9   & 9.67  & 1.21\\
NGC~4235        & 1.05  & 1.5 & 12.26 & 1.03 & 58\% & Extended\\
                &       & 3   & 11.44 & 0.76\\
                &       & 6   & 10.77 & 0.46\\
                &       & 9   & 10.37 & 0.24\\
NGC~4253/Mkn~766 & 0.69 & 1.5 & 11.37 & 1.87 & 90\% & Compact\\
                &       & 3   & 10.96 & 1.75 \\
                &       & 6   & 10.70 & 1.60\\
                &       & 9   & 10.55 & 1.50\\
NGC~4395        & \nodata  & 6   & 14.00 & 1.29 &\nodata&  Blind Offsets\\ 
NGC~4579        & 1.02   & 1.5 & 11.28 & 0.49 & 45\% & Extended\\
                &       & 3   & 10.32 & 0.33\\
                &       & 6   & 9.57  & 0.22\\
                &       & 9   & 9.17  & 0.15\\
NGC~5033        & 0.98  & 1.5 & 11.59 & 0.66 & 35\% & Extended\\
                &       & 3   & 10.59 & 0.42 \\
                &       & 6   & 9.77  & 0.26\\
                &       & 9   & 9.31  & 0.19\\
\tablebreak
NGC~5256/Mkn~266SW & -- & 1.5 & 13.21 & 0.93 & 16\% & Extended \\
                &       & 3   & 12.30 & 0.98\\ 
                &       &6    & 11.71 & 0.94 \\
                &       & 9   & 11.44 & \nodata \\
NGC~5256/Mkn~266NE$^*$ & -- & 1.5 & 13.51 & 0.58 & \nodata & Extended\\
                &       & 3   & 12.67 & 0.65 \\        
                &       & 6   & 12.08 & 0.67 \\
                &       & 9   & 11.77 & \nodata\\
NGC~5273        & 1.6?  & 1.5 & 12.98 & 0.64 & 38\% & Extended\\
                &       & 3   & 11.92 & 0.48\\ 
                &       & 6   & 11.17 & 0.27 \\
                &       & 9   & 10.83 & \nodata\\
NGC~5283/Mkn~270 & --   & 6   & 11.26 & 0.22 & \nodata & Blind Offsets\\
Mkn~461         & --    & 6   & $L=11.97$& & \nodata & Blind Offsets\\
NGC~5347        & 0.75  & 1.5 & 12.80 & 2.09 & 65\% & Compact\\
                &       & 3   & 12.09 & 1.80 \\
                &       & 6   & 11.62 & 1.51\\
                &       & 9   & 11.34 & 1.30\\ 
NGC~5548        & 0.84  & 1.5 & 11.26 & 1.93 & 65\% & Compact\\
                &       & 3   & 10.77 & 1.82 \\
                &       & 6   & 10.46 & 1.66\\
                &       & 9   & 10.29 & 1.53\\
NGC~5674        & 0.83  & 1.5 & 12.69 & 1.41 & 57\% & Compact\\
                &       & 3   & 12.09 & 1.30 \\
                &       & 6   & 11.66 & 1.12 \\
                &       & 9   & 11.43 & 1.03\\
Mkn~841         & 0.90   & 1.5 & 12.88 & 2.10 & 57\% & Compact\\
                &       & 3   & 12.34 & 2.04\\
                &       & 6   & 12.10 & 1.97\\
                &       & 9   & 11.98 & 1.92\\
NGC~5929        & 1.50   & 1.5 & 13.07 & 0.43 & 16\% & Extended\\
                &       & 3   & 12.13 & 0.27 \\
                &       & 6   & 11.45 & 0.04\\
NGC~5930$^*$   & 1.95  & 1.5 & 12.37 & 0.63 &      & Extended \\
                &       & 3   & 11.41 & 0.63\\
                &       & 6   & 10.86 & 0.43\\
                &       & 9   & 10.57 & 0.28\\
NGC~5940        & \nodata    & 1.5 & 13.95 & \nodata & \nodata & Not detected\\
                &       & 3   & 13.11 \\
                &       & 6   & 12.52 \\
                &       & 9   & 12.18\\
Mkn~1513        & 0.77  & 1.5 & 11.23 & 1.68 & 75\% & Compact\\
                &       & 3   & 10.89 & 1.67\\
                &       & 6   & 10.78 & 1.66\\
UGC~12138       & 0.86  & 1.5 & 12.39 & 1.36 & 54\% & Compact\\
                &       & 3   & 11.81 & 1.21 \\
                &       & 6   & 11.46 & 1.07\\
                &       & 9   & 11.29 & 0.98\\
\tablebreak
NGC~7469        & 0.75  & 1.5 & 11.17 & 1.96 & 71\% & Compact nucleus + \\
                &       & 3   & 10.41 & 1.64 &      & Extended\\
                &       & 6   & 9.90  & 1.42\\
                &       & 9   & 9.73  & 1.34\\
NGC~7479        & 0.75  & 1.5 & 12.78 & 1.59 & 61\% & Compact nucleus + \\
                &       & 3   & 11.95 & 1.22 & & Extended\\
                &       & 6   & 11.19 & 0.83 &\\
                &       & 9   & 10.74 & 0.64\\ 
NGC~7603/Mkn~530 & 0.69 & 1.5 & 10.85 & 1.59 & 100\% & Compact\\
                &       & 3   & 10.45 & 1.48\\
                &       & 6   & 10.21 & 1.37 \\
                &       & 9   & 10.07 & 1.27\\
NGC~7682        &\nodata & 6  & 11.92 & 0.05 & \nodata & Blind Offsets\\
NGC~7743        & 1.00   & 1.5 & 11.93 & 0.16 & $<45\%$ & Extended\\
                &       &  3  & 11.22 & 0.19 \\
                &       &  6  & 10.67 & 0.20\\
                &       &  9  & 10.38 & 0.25\\ 
\enddata
\tablecomments{Column~(1): Galaxy. Column~(2): Size (FWHM) of the nucleus 
as measured 
from the $L$-band image. Column~(3): Diameter in arcsec of the 
circular apertures used for the photometry. 
Column~(4): $K^\prime$-band magnitude. 
Column~(5): measured $K^\prime-L$ color. Column~(6):
Fraction of 
unresolved emission within a 3\arcsec-diameter apeture in the $L$-band.  
Column~(7): Morphology of the $L$-band emission.\\
$^*$ NGC~5930 and Mkn~266NE are not part of the CfA sample of
Seyfert galaxies. }
\end{deluxetable}

\begin{deluxetable}{ccccc}
\footnotesize
\tablewidth{15cm}
\tablecaption{Galaxies with circumnuclear radio and mid-infrared emission.}
\tablehead{
\multicolumn{1}{l}{Galaxy}      &
\colhead{6\,cm $\frac{f_{0.3\arcsec}}{f_{15\arcsec}}$}          &
\colhead{$10\,\micron$ $\frac{f_{\rm ground}}{f_{\rm ISO}}$ }          &
\colhead{Extended $L$}          &
\colhead{Extended {\it ISO}}\\
\colhead{(1)}                        & 
\colhead{(2)}                        & 
\colhead{(3)}                        & 
\colhead{(4)}                        & 
\colhead{(5)}} 
\startdata
Mkn~334  & 0.12 & 1.18 & no & no\\
NGC~1144 & 0.07 & \nodata & yes & yes (at $6.75\,\micron$) \\
NGC~3982 & 0.03 & 0.34 & \nodata & yes\\
NGC~4051 & 0.07 & 0.77 & no & no\\
NGC~4388 & 0.08 & 1.90 & no & yes + bright point source\\
NGC~5033 & 0.15 & 0.19 & yes & yes\\
NGC~5256/Mkn~266SW  & 0.10 & 0.33 & yes & yes + bright point source\\
Mkn~841  & 0.14 & 1.28 & no & no\\
NGC~5929 & 0.40 & 0.41 & yes & yes\\
NGC~5940 & \nodata & 0.45 & \nodata & yes\\
\enddata
\tablecomments{Column~(1): Galaxy. Column~(2): Ratio of 
the nuclear (0.3\arcsec) to extended (15\arcsec)
emission at 6\,cm. Column~(3): Ratio of ground-based 
to {\it ISO} fluxes at $10\,\micron$. 
Column~(4) and Column~(5) indicate whether there is
evidence for significant extended emission from the IRTF 
$L$-band images and {\it ISO} $9.63\,\micron$ images, 
respectively.}
\end{deluxetable}

\begin{deluxetable}{lcccccccccc}
\scriptsize
\tablecaption{Nuclear (unresolved) optical and infrared fluxes for Seyfert 
$1.8-2$ Galaxies in the extended CfA sample, and fitted spectral indices.}
\tablehead{
\multicolumn{1}{l}{Galaxy}      &
\colhead{$f_{\rm opt}$}          &
\colhead{$f_{1.1}$}          &
\colhead{$f_{1.6}$}          &
\colhead{$f_{2.1}$}          &
\colhead{$f_{3.5}$}          &
\colhead{$f_{6.75}$}         &
\colhead{$f_{10}$}         &
\colhead{$f_{16}$}         &
\colhead{$\alpha_{\rm opt-IR}$}         &
\colhead{$\alpha_{\rm IR}$}    \\
\colhead{(1)}                        & 
\colhead{(2)}                        & 
\colhead{(3)}                        & 
\colhead{(4)}                        & 
\colhead{(5)}                        & 
\colhead{(6)}                        &
\colhead{(7)}                        &
\colhead{(9)}                        &
\colhead{(9)}                        &
\colhead{(10)}                       &
\colhead{(11)}
} 
\startdata
 
 Mkn~334  &
  $< 3.8$& 
  2.33 & 
  6.83 & 
  13.8 & 
  31.7 & 
  131. & 
  148. & 
 $< 390.$ & 
$1.80\pm0.11$ &
$1.80\pm0.11$
  \\
 Mkn~573  &
 $<   0.097$ & 
  0.15 & 
  0.54 & 
  3.2 & 
  18.8 & 
  \nodata & 
  $167.^*$ & 
 $<   550.$ & 
$2.98\pm0.20$&
$2.91\pm0.23$  \\
 UM~146/UGC~1395   &
  0.090 & 
  0.35 & 
  0.82 & 
  3.0 & 
  4.4 & 
  15. & 
  $15.^*$ & 
 $<   250.$ & 
$1.98\pm0.18$&  
$1.76\pm0.18$
\\
NGC~1068 & 
\nodata  & 
9.8 &
98. &
445. &
3080.&
13600.&
19600. &
$<46900.$ &
$3.02\pm0.31$ &
$3.02\pm0.31$
\\

 NGC~1144 &
 $<   0.020$ & 
 $<   0.11$ & 
  0.08 & 
  \nodata & 
  3.4 & 
 $<   120.$ & 
  $<158.^*$ & 
 $<   460.$ & 
$3.3$ &
$3.3$
 \\
\\
NGC~3079 & \nodata & \nodata & $<0.1$ & \nodata & $<61$ & \nodata & 
$92.^*$ & $<2070$
& $3.7$ & $3.7$\\
NGC~3185 & \nodata & \nodata & \nodata & \nodata & \nodata & \nodata & 
$20.^*$ & \nodata & 
\nodata & \nodata\\
 NGC~3362 &
 $<   0.031$ & 
  0.04 & 
  0.05 & 
  \nodata & 
 $<   2.2$ & 
  \nodata & 
  $12.^*$ & 
 $<   210.$ &
$2.75\pm0.20$&
$2.75\pm0.20$

  \\
 UGC~6100 &
 $<   0.053$ & 
 $<   0.17$ & 
 $<   0.15$ & 
  \nodata & 
 $<   4.6$ & 
  42. & 
  46. & 
 $<   130.$ & 
$3.4$ &
3.4
  \\
 NGC~3786/Mkn~744 &
  2.2 & 
  1.75 & 
  3.25 & 
  $<11.9$ & 
  17.7 & 
  \nodata & 
  $59.^*$ & 
 $<   190.$ &
$1.22\pm0.19$ &
$1.65\pm0.09$ 
  \\
\\
 NGC~3982 &
  0.009 & 
 $<   0.27$ & 
  0.34 & 
  \nodata & 
 $<   7.2$ & 
  $<62.$ & 
   $19.^*$ & 
 $<   430.$ &
$2.77\pm0.19$ &
$2.4$ 
  \\
NGC~4258 & $<0.019$ & $<0.5$ & 0.9 & 4.0 & 20. & \nodata & $100.^*$ & \nodata & 
$2.85\pm0.30$ & $2.39\pm0.21$\\
 NGC~4388 &
  \nodata & 
  0.06 & 
  0.71 & 
  \nodata & 
  39.9 & 
  265. & 
  267. & 
 $<   1550.$ &
$3.51\pm0.42$ &
$3.51\pm0.42$ 
  \\
NGC~4501 & \nodata & \nodata & \nodata& \nodata & \nodata & $<35$ & $17.^*$ & \nodata &
\nodata & \nodata\\
NGC~4579 & 0.202 & \nodata & \nodata & $<14$ & 12.5 & 104. & 113. & 
& $2.30\pm0.13$ & 
$2.35\pm0.46$ \\
\\
 NGC~5033 &
  0.89 & 
  0.61 & 
  3.22 & 
 $<   8.0$ & 
  8.6 & 
  $<179.$ & 
  $24.^*$ & 
 $<   1340.$ &
$1.41\pm0.20$ &
$1.67\pm0.22$ 
  \\
NGC~5194 & \nodata & $<0.14$ & 0.19 & 0.38 & $<34.$ & \nodata & 110. & \nodata&
$3.78\pm0.11$& $3.78\pm0.11$\\
 UGC~8621/1335+39 &
  \nodata & 
  \nodata & 
  \nodata & 
  \nodata & 
  \nodata & 
  49. & 
  49. & 
 $<   130.$ & 
\nodata&
\nodata
  \\
 NGC~5252 &
  0.070 & 
  0.20 & 
  0.70 & 
  1.0 & 
  11.1 & 
  15. & 
  27. & 
  $<45.$ &
$2.03\pm 0.20$ &
$1.94\pm0.26$ 
  \\
NGC~5256/Mkn~266SW &
  \nodata & 
 $<   0.08$ & 
 $<   0.15$ & 
  \nodata & 
  2.4 & 
  $<125.$ & 
  $30.^*$ & 
 $<   330.$ & 
$3.0$&
$3.0$ 
\\
\\
 NGC~5283/Mkn~270 &
 $<   0.013$ & 
 $<   0.43$ & 
 $<   0.30$ & 
  \nodata & 
 $<   11.0$ & 
 $<   11.$ & 
  $17.^*$ & 
 $<   60.$ & 
2.7&
2.7
  \\
 NGC~5273 &
  0.27 & 
  0.68 & 
  1.67 & 
  \nodata & 
  3.0 & 
  23. & 
  25. & 
 $<   90.$ &
$1.63\pm0.11$ &
$1.66\pm0.15$ 
  \\
 Mkn~461  &
  \nodata & 
 $<   0.45$ & 
 $<   0.46$ & 
  \nodata & 
 $<   4.7$ & 
 $<   12.$ & 
  $12.^*$ & 
  \nodata & 
1.8 &
1.8
  \\
 NGC~5347 &
  \nodata & 
  0.22 & 
  0.97 & 
  4.0 & 
  14.3 & 
  \nodata & 
  $208.^*$ & 
  \nodata & 
$3.06\pm0.18$&
$3.06\pm0.18$
  \\
 Mkn~471  &
  0.15 & 
  0.18 & 
  0.32 & 
  \nodata & 
  3.0 & 
  \nodata & 
  $14.^*$ & 
  \nodata & 
$1.82\pm0.18$ &
$2.08\pm0.14$
  \\
\\
 NGC~5674 &
  0.11 & 
  0.90 & 
  2.62 & 
  4.2 & 
  7.5 & 
  38. & 
  43. & 
  \nodata & 
$2.08\pm0.16$&
$1.78\pm0.11$
  \\
 NGC~5695/Mkn~686  &
 $<   0.039$ & 
 $<   0.17$ & 
 $<   0.25$ & 
  \nodata & 
  \nodata & 
  \nodata & 
 $<   10.^*$ & 
  \nodata & 
\nodata &
\nodata
  \\
 NGC~5929 &
 $<   0.023$ & 
 $<   0.17$ & 
 $<   0.25$ & 
  \nodata & 
  1.3 & 
  $<41.$ & 
  $24.^*$ & 
 $<   550.$ &
$2.9$ &
$2.9$
  \\
 UGC~12138&
  $<2.4$ & 
  1.86 & 
  2.59 & 
  5.3 & 
  9.0 & 
  40. & 
  54. & 
 $<   250.$ & 
$1.64\pm0.08$ &
$1.64\pm0.08$  \\
NGC~7479 & $<0.006$ & \nodata & 0.24 & $<3.$ & 9.0 & 150. & $263.^*$ 
& \nodata & 
$3.85\pm0.24$ & $3.58\pm 0.30$\\
\\
 NGC~7674/Mkn~533 &
  \nodata & 
  1.25 & 
  5.0 & 
  12.3 & 
  53. & 
  259. & 
  344. & 
 $<   860.$ & 
$2.36\pm 0.18$&
$2.36\pm 0.18$
  \\
 NGC~7682 &
 $<   0.04$ & 
 $<   0.10$ & 
 $<   0.13$ & 
  \nodata & 
 $<   4.8$ & 
  10. & 
  $12.^*$ & 
 $<   290.$ & 
3.0 &
3.1\\
NGC~7743 & \nodata & \nodata& $<0.5$& $<8.$ & $<5.$& 
\nodata & \nodata & \nodata & \nodata & \nodata \\

\enddata
\tablecomments{
This table contains all the optically classified 
Seyfert $1.8-2$ galaxies in the 'extended' CfA sample.
The measured unresolved flux densities in Columns~(2) through 
(9) are listed in mJy. Columns~(10) and (11) are the fitted 
spectral indices ($f_\nu \propto \nu^{-\alpha}$) in the 
optical-infrared ($\alpha_{\rm opt-IR}$, 
spectral range $\simeq 0.4-16\,\micron$) and infrared
($\alpha_{\rm IR}$, spectral range $1-16\,\micron$), respectively.
If  no optical measurement is available, we give 
the same value for $\alpha_{\rm opt-IR}$ and $\alpha_{\rm IR}$.
When the errors are not quoted for the fitted spectral 
indices it means that the formal errors are infinity (those cases with only
two or fewer detections). \\
Column~(9) $^*$ indicates small aperture ground-based $N$-band measurement (see discussion 
in Section~4.3). The ground-based $N$-band fluxes
(derived to give monochromatic fluxes of Rayleigh-Jeans
spectra) need to be corrected to $\nu \ f_{\nu} = {\rm constant}$ by 
multiplying them by a factor of 1.22 (see also Edelson et al. 1987, and
Section~5).\\
Additional nuclear fluxes at $4.8\,\micron$ (ground-based) --- 
Mkn~573: 41.3\,mJy; NGC~1068: 
8245\,mJy ; NGC~5252: 15.7\,mJy; NGC~7674/Mkn~533: 108.6\,mJy 
(Alonso-Herrero et al. 2001). Additional nuclear fluxes at 
$12\,\micron$ ({\it ISO}) --- NGC~3185: 20\,mJy;  NGC~5194: 400\,mJy;
NGC~5695/Mkn~686: $<55\,$mJy. Additional nuclear fluxes at $12\,\micron$ (ground-based):
NGC~4258: 165\,mJy.
Additional nuclear fluxes at $14.3\,\micron$ ({\it ISO}) --- NGC~7479: 520\,mJy. 
Additional nuclear fluxes at $17.9\,\micron$ (ground-based) --- NGC~4258: 300\,mJy.}
\end{deluxetable}

\begin{deluxetable}{lcccccccccc}
\scriptsize
\tablecaption{Nuclear (unresolved) optical and infrared fluxes for Seyfert 
$1-1.5$ Galaxies in the extended CfA sample.}
\tablehead{
\multicolumn{1}{l}{Galaxy}      &
\colhead{$f_{\rm opt}$}          &
\colhead{$f_{1.1}$}          &
\colhead{$f_{1.6}$}          &
\colhead{$f_{2.1}$}          &
\colhead{$f_{3.5}$}          &
\colhead{$f_{6.75}$}         &
\colhead{$f_{10}$}         &
\colhead{$f_{16}$}         &
\colhead{$\alpha_{\rm opt-IR}$}         &
\colhead{$\alpha_{\rm IR}$}    \\
\colhead{(1)}                        & 
\colhead{(2)}                        & 
\colhead{(3)}                        & 
\colhead{(4)}                        & 
\colhead{(5)}                        & 
\colhead{(6)}                        &
\colhead{(7)}                        &
\colhead{(9)}                        &
\colhead{(9)}                        &
\colhead{(10)}                       &
\colhead{(11)}             
} 
\startdata

 Mkn~335  &
  4.0 & 
  \nodata & 
  \nodata & 
  32.8 & 
  78.2 & 
  155. & 
  203. & 
  $<240.$ & 
$1.31\pm0.18$ &
$0.93\pm0.15$
  \\
 UGC~524/0048+29  &
  \nodata & 
  \nodata & 
  \nodata & 
  \nodata & 
  \nodata & 
  \nodata & 
  $41.^*$ & 
  \nodata & 
\nodata&
\nodata
  \\
 Mkn~993  &
  0.039 & 
  \nodata & 
  \nodata & 
  $<3.1$ & 
 $<   11.$ & 
  \nodata & 
  $18.^*$ & 
  $<130.$ &
$2.2$ &
1.6
  \\
 I~Zw~1    &
  2.20 & 
  \nodata & 
  \nodata & 
  \nodata & 
  110. & 
  \nodata & 
  $310.^*$ & 
  $<640.$ & 
$1.64\pm0.14$ &
$1.07\pm0.10$
  \\
 Mkn~590  &
  0.80 & 
  \nodata & 
  \nodata & 
  9.3 & 
  38.0 & 
  144. & 
  198. & 
  $<460.$ & 
$2.08\pm0.14$ &
$1.81\pm0.17$
  \\
\\
 Mkn~1243 &
  \nodata & 
  \nodata & 
  \nodata & 
  \nodata & 
  \nodata & 
  \nodata & 
  $16.^*$ & 
  $<210.$ & 
\nodata &
\nodata
  \\
NGC~3031 & 
1.39 & 
\nodata &
13.4 &
$<202.$ &
$<161.$ &
\nodata & 
$86.^*$ & 
\nodata& 
$1.42\pm0.15$ &
$1.1$
\\
 NGC~3227 &
  2.70 & 
  \nodata & 
  10.6 & 
  22.6 & 
  46.7 & 
  294. & 
  382. & 
  $<1220.$ &
$1.80\pm0.10$ &
$1.97\pm0.11$ 
  \\
 NGC~3516 &
  2.20 & 
  \nodata & 
  18.1 & 
  \nodata & 
  117. & 
  264. & 
  369. & 
  $<770.$ & 
$1.71\pm0.11$ &
$1.51\pm0.16$
  \\
 NGC~4051 &
  1.60 & 
  \nodata & 
  \nodata & 
  38.2 & 
  73.5 & 
  265. & 
  411. & 
  $<1440$. & 
$1.91\pm0.08$ &
$1.62\pm0.07$
  \\
\\
 NGC~4151 &
  51.4 & 
  69.0 & 
  103.6 & 
  177.5 & 
  325. & 
  \nodata & 
  $1400.^*$ & 
  $<4120.$ & 
$1.26\pm0.08$&
$1.44\pm0.03$
  \\
 NGC~4235 &
  0.44 & 
  \nodata & 
  3.69 & 
  $<6.6$ & 
  9.0 & 
  \nodata & 
  $34.^*$ & 
  $<170.$ & 
$1.51\pm 0.09$ &
$1.30\pm0.04$
  \\
 NGC~4253/Mkn~766 &
  3.90 & 
  \nodata & 
  20.0 & 
  20.3 & 
  58.0 & 
  177. & 
  274. & 
  $<730.$ & 
$1.65\pm0.07$ &
$1.56\pm0.08$
  \\
 Mkn~205  &
  2.10 & 
  \nodata & 
  \nodata & 
  \nodata & 
  13. & 
  30. & 
  $58.^*$ & 
  $<190.$ & 
$1.34\pm0.06$ &
$1.53\pm0.10$ \\
 NGC~4395 &
  0.17 & 
  \nodata & 
  0.85 & 
  $<1.1$ & 
 $<   2.4$ & 
  \nodata & 
  $12.^*$ & 
  \nodata &
$1.29\pm0.10$ &
1.5 
  \\
\\
 Mkn~231  &
  16.3 & 
  \nodata & 
  80.1 & 
  \nodata & 
  326. & 
  \nodata & 
  $1420.^*$ & 
  $<3960.$ & 
$1.73\pm0.07$ &
$1.59\pm0.06$
  \\
 Mkn~279  &
  4.2 & 
  \nodata & 
  \nodata & 
  \nodata & 
  32. & 
  102. & 
  146. & 
  $<290.$ & 
$1.14\pm0.06$ &
$1.38\pm0.11$
  \\
 NGC~5548 &
  4.3 & 
  \nodata & 
  15.0 & 
  31.6 & 
  46.2 & 
  177. & 
  268. & 
  $<440.$ & 
$1.52\pm0.09$ &
$1.43\pm0.10$
  \\
 Mkn~817  &
  2.2 & 
  \nodata & 
  \nodata & 
  \nodata & 
  55. & 
  142. & 
  259. & 
  $<490.$ & 
$1.78\pm0.11$ &
$1.43\pm0.05$
  \\
 Mkn~841  &
  1.8 & 
  \nodata & 
  \nodata & 
  4.1 & 
  12.0 & 
  74. & 
  133. & 
  $<340.$ & 
$1.75\pm0.17$ &
$2.18\pm0.12$
  \\
\\
 NGC~5940 &
  0.46 & 
  \nodata & 
  0.93 & 
  $<1.9$ & 
  \nodata & 
  $<46.$ & 
  $26^*$ & 
  \nodata & 
$1.54\pm0.18$ &
$1.8$
  \\
 Mkn~1513 &
  \nodata & 
  \nodata & 
  \nodata & 
  19.7 & 
  44.6 & 
  85. & 
  $140.^*$ & 
  \nodata & 
$1.08\pm0.12$ &
$1.08\pm0.12$
  \\
 NGC~6104 &
  0.12 & 
  \nodata & 
  0.13 & 
  \nodata & 
  \nodata & 
  \nodata & 
  $12.^*$ & 
  $<120.$ & 
$1.76\pm0.35$ &
$2.5$
  \\
 NGC~7469 &
  6.4 & 
  16.2 & 
  39.0 & 
  67.8 & 
  90. & 
  433. & 
  600. & 
  $<1290.$ & 
$1.67\pm0.09$ &
$1.57\pm0.09$
  \\
 NGC~7603/Mkn~530 &
  0.53 & 
  \nodata & 
  \nodata & 
  21. & 
  74.0 & 
  138. & 
  $176.^*$ & 
  \nodata & 
$2.29\pm0.32$ &
$1.35\pm0.19$
  \\

\enddata
\tablecomments{This table contains all the optically classified 
Seyfert $1-1.5$ galaxies in the 'extended' CfA sample. 
The measured unresolved flux densities in Columns~(2) through 
(9) are listed in mJy. Columns~(10) and (11) are the fitted 
spectral indices ($f_\nu \propto \nu^{-\alpha}$) in the 
optical-infrared ($\alpha_{\rm opt-IR}$, 
spectral range $\simeq 0.4-16\,\micron$) and infrared
($\alpha_{\rm IR}$, spectral range $1-16\,\micron$), respectively.
If no optical measurement is available, we give 
the same value for $\alpha_{\rm opt-IR}$ and $\alpha_{\rm IR}$.
When the errors are not quoted for the fitted spectral 
indices it means that the formal errors are infinity (those cases with only
two or fewer detections). \\
Column~(9) $^*$ indicates small aperture ground-based $N$-band measurement (see discussion 
in Section~4.3). The ground-based $N$-band fluxes
(derived to give monochromatic fluxes of Rayleigh-Jeans
spectra) need to be corrected to $\nu \ f_{\nu} = {\rm constant}$ by 
multiplying them by a factor of 1.22 (see also Edelson et al. 1987, and
Section~5).\\
Additional fluxes at $5.5\,\micron$ ({\it ISO}) --- I~Zw~1: 173\,mJy, NGC~4151: 882\,mJy,
Mkn~231: 606\,mJy. 
Additional fluxes at $7.7\,\micron$ ({\it ISO}) --- I~Zw~1: 221\,mJy, NGC~4151: 1090\,mJy,
Mkn~231: 919\,mJy. Additional fluxes at $12\,\micron$ ({\it ISO}) --- 
NGC~4395: 7.6\,mJy. Additional fluxes at $14.3\,\micron$ ({\it ISO}) --- 
Mkn~1513: 160\,mJy.}
\end{deluxetable}

\begin{deluxetable}{cccccc}
\tablewidth{9cm}
\tablecaption{Parameters of Efstathiou \& Rowan-Robinson (1995) 
and Efstathiou et al. (1995) torus models.}
\tablehead{
\multicolumn{1}{l}{Model}    &
\colhead{$\Theta_{\rm c}$}   &
\colhead{$\tau_{\rm UV}$}    &
\colhead{$r_1/r_2$}          &
\colhead{$h/r_2$}            &
\colhead{$T$}  
\\
\colhead{(1)}                        & 
\colhead{(2)}                        & 
\colhead{(3)}                        & 
\colhead{(4)}                        &
\colhead{(5)}                        &
\colhead{(6)}} 

\startdata

Torus 1 & $45\arcdeg$ & 1000 & 0.05 & 0.5& 1000\\
Torus 2 & $45\arcdeg$ & 500 & 0.05 & 0.5 & 1000\\
Torus 3 & $45\arcdeg$ & 250 & 0.05 & 0.5 & 1000\\
Torus 4 & $30\arcdeg$ & 1200 & 0.01 & 0.1 & 900\\
Torus+Cone & $30\arcdeg$ & 1200 & 0.01 & 0.1 & 900\\

\enddata
\tablecomments{Column~(2): $\Theta_{\rm c}$ is the half opening 
angle of the torus. Column~(3): $\tau_{\rm UV}$ 
is the equatorial UV optical depth in magnitudes. 
Column~(4): $r_1/r_2$ is ratio of the inner and outer torus radii.
Column~(5): $h/r_2$ is the ratio of the height
scale and the outer radius. Column~(6): $T$ (in K) 
is the dust sublimation temperature.}
\end{deluxetable}

\begin{figure*}
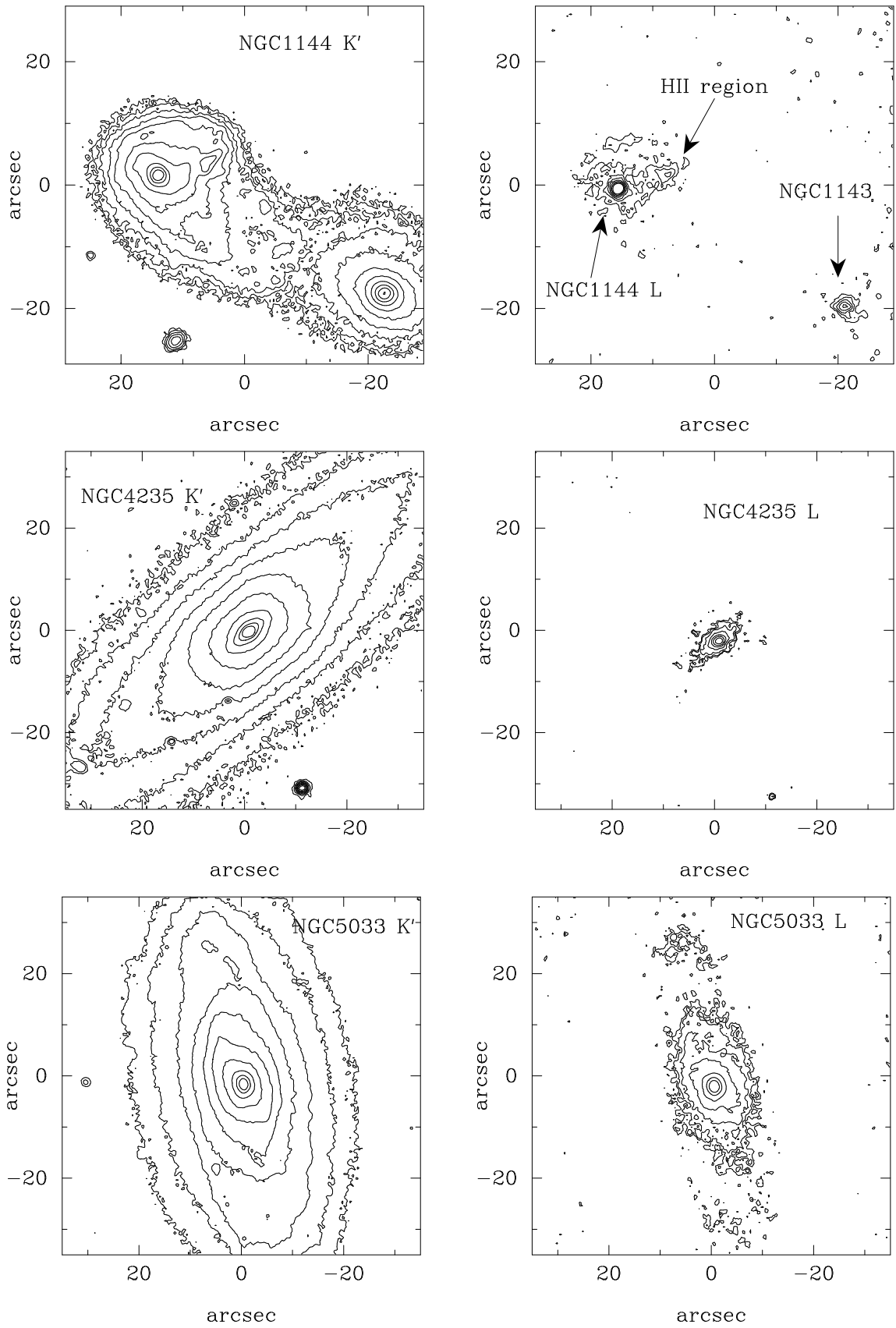

\figurenum{1}
\plotfiddle{figure1a.ps}{425pt}{-90}{60}{60}{-250}{550}
\plotfiddle{figure1b.ps}{425pt}{-90}{60}{60}{-250}{770}
\plotfiddle{figure1c.ps}{425pt}{-90}{60}{60}{-250}{990}
\vspace{-23cm}
\caption{Contours plots of selected galaxies with 
extended emission in the $L$-band (right panels). We 
show the  $K^\prime$-band contour plots for comparison (left panels).
The orientation of the images is south up, east to the left.}
\end{figure*}

\begin{figure*}
\figurenum{2}
\plotfiddle{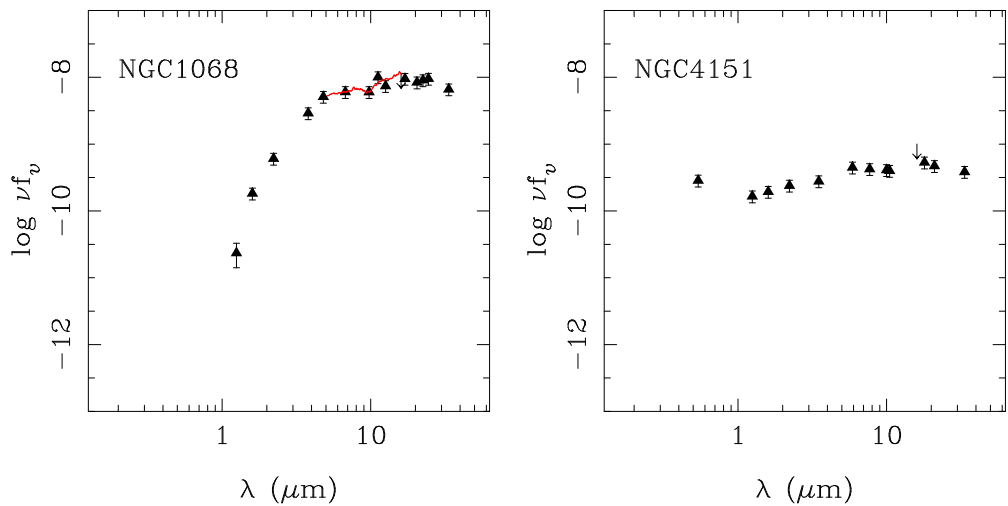}{425pt}{-90}{60}{60}{-200}{500}
\vspace{-5.5cm}
\caption{{\it Left panel:} Non-stellar SED of NGC~1068 in units of 
erg\,cm$^{-2}$\,s$^{-1}$. The {\it ISO} spectrum 
of the AGN (nuclear) component of NGC~1068 
is from Le Floc'h et al. (2001).
The additional mid-infrared data points
at 12.6, 17, 22.5, 24.5, and $33.5\,\mu$m
are ground-based small aperture observations from Rieke \& Low (1975a); 
data points at 11.2 and $20.5\,\micron$ are from Alloin et 
al. (2000).
{\it Right panel:} Non-stellar SED of NGC~4151 in units of 
erg\,cm$^{-2}$\,s$^{-1}$. The additional mid-infrared data points
at 10 and $18\,\mu$m (unresolved emission) are from Radomski et 
al. (2002), at $21\,\mu$m is from Rieke \& Low (1972) and 
at $33.5\,\mu$m from Rieke \& Low (1975b).}
\end{figure*}

\begin{figure*}
\figurenum{3a}
\plotfiddle{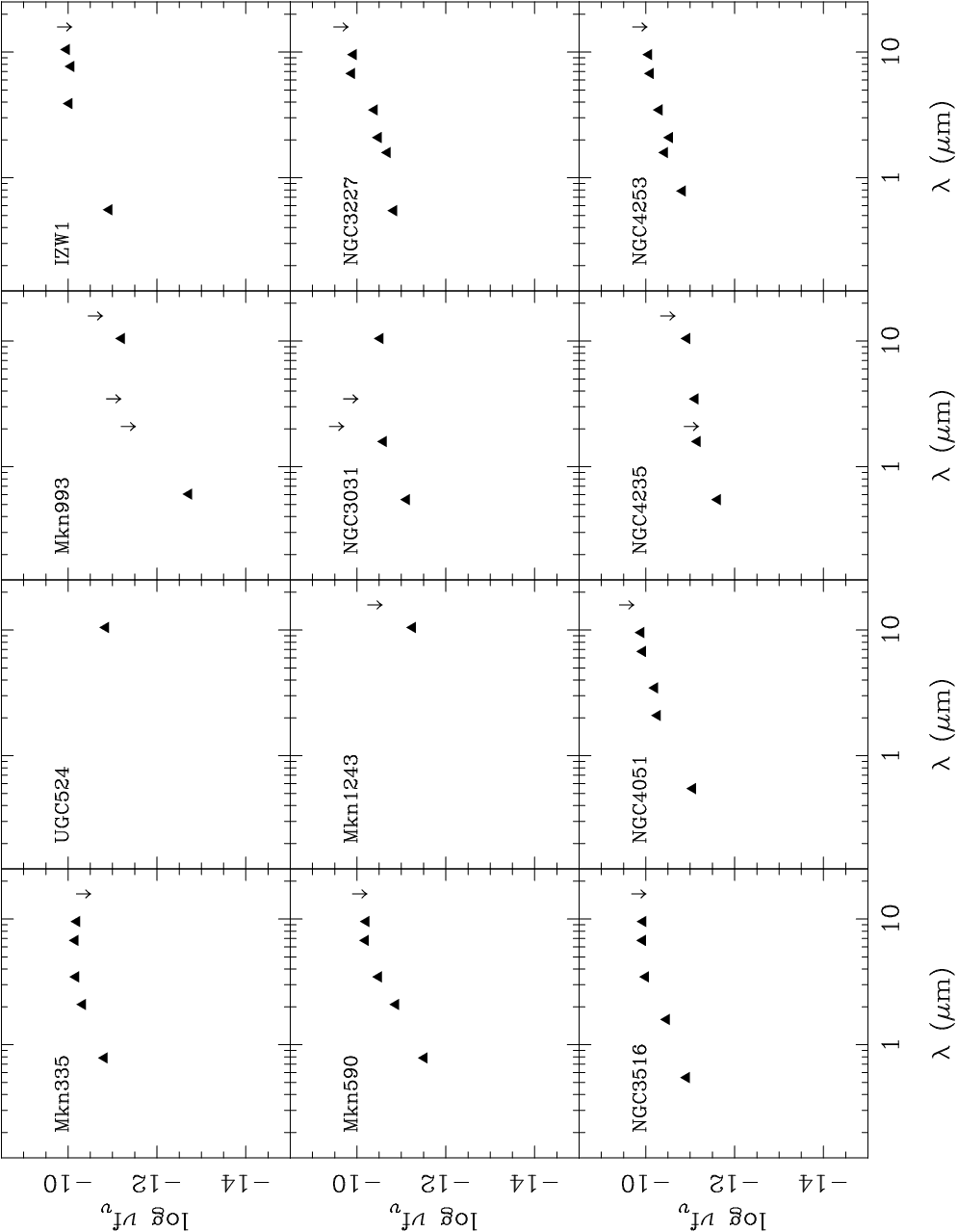}{425pt}{-90}{70}{70}{-250}{400}
\caption{Observed non-stellar 
SEDs ($\nu\,f_{\nu}$)
of optically classified  type 1 ($1-1.5$) CfA Seyferts  in 
units of erg\,cm$^{-2}$\,s$^{-1}$.}
\end{figure*}

\begin{figure*}
\figurenum{3b}
\plotfiddle{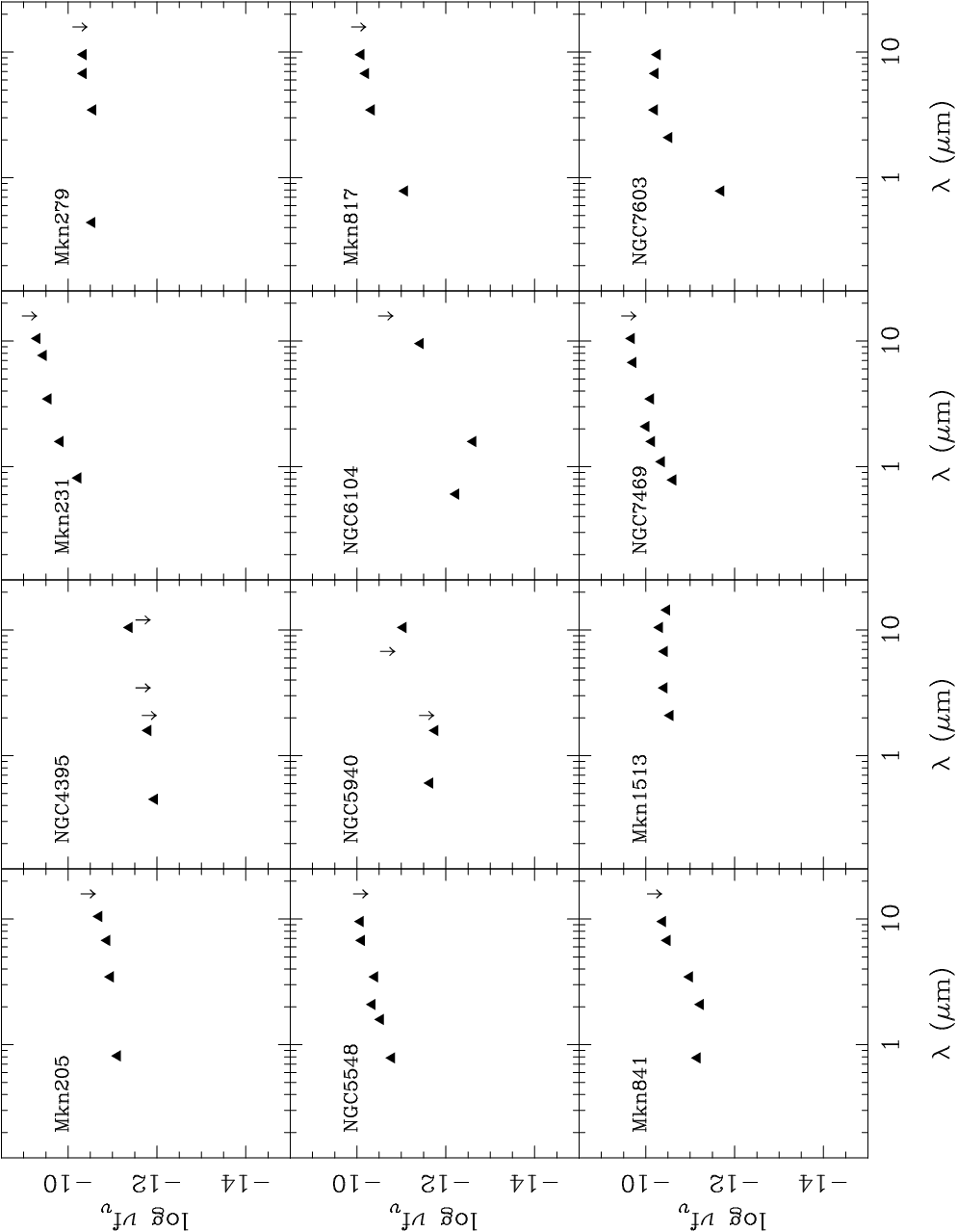}{425pt}{-90}{70}{70}{-250}{400}
\caption{Continued.}
\end{figure*}

\begin{figure*}
\figurenum{4a}
\plotfiddle{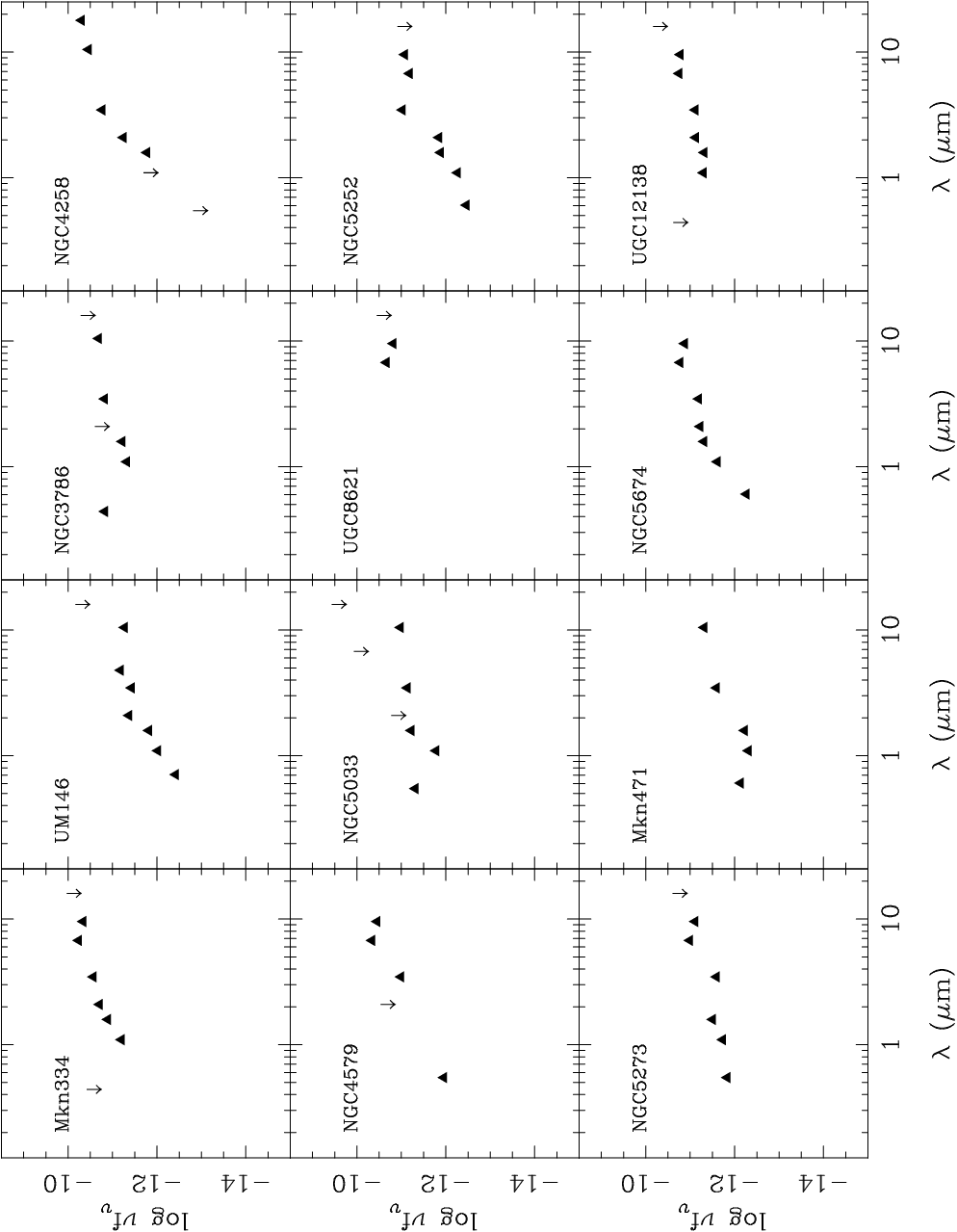}{425pt}{-90}{70}{70}{-250}{400}
\caption{Observed non-stellar 
SEDs ($\nu\,f_{\nu}$) of 
optically classified intermediate type ($1.8-1.9$) CfA Seyferts in 
units of erg\,cm$^{-2}$\,s$^{-1}$.}
\end{figure*}

\begin{figure*}
\figurenum{4b}
\plotfiddle{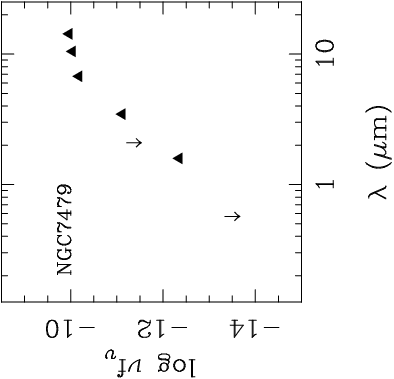}{425pt}{-90}{70}{70}{-250}{400}
\caption{Continued.}
\end{figure*}

\begin{figure*}
\figurenum{5a}
\plotfiddle{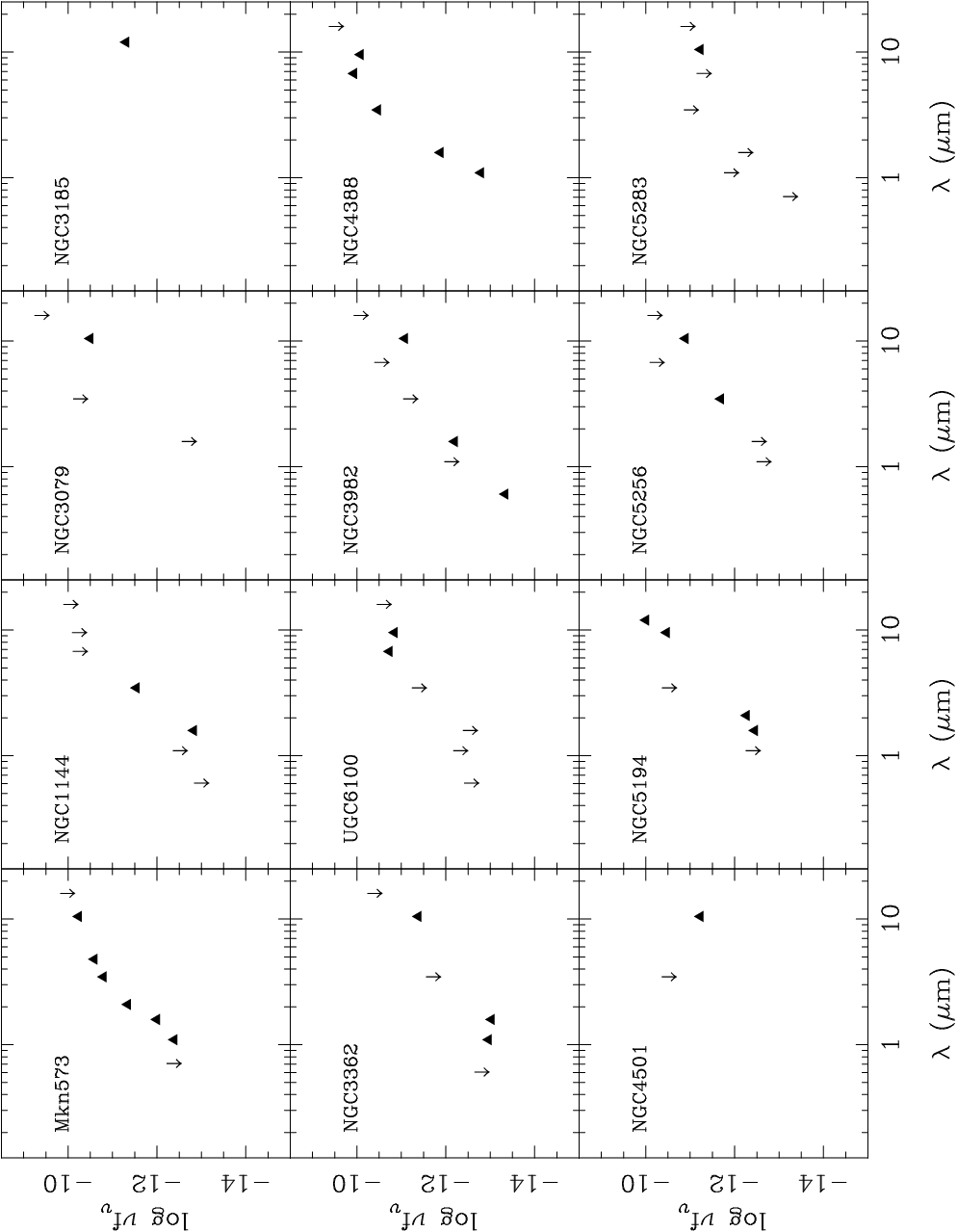}{425pt}{-90}{70}{70}{-250}{400}
\caption{Observed non-stellar 
SEDs ($\nu\,f_{\nu}$) of optically classified type 2 CfA Seyferts in 
units of erg\,cm$^{-2}$\,s$^{-1}$.
}
\end{figure*}

\begin{figure*}
\figurenum{5b}
\plotfiddle{figure5b.ps}{425pt}{-90}{70}{70}{-250}{400}
\caption{Continued.}
\end{figure*}

\begin{figure*}
\figurenum{6}
\plotfiddle{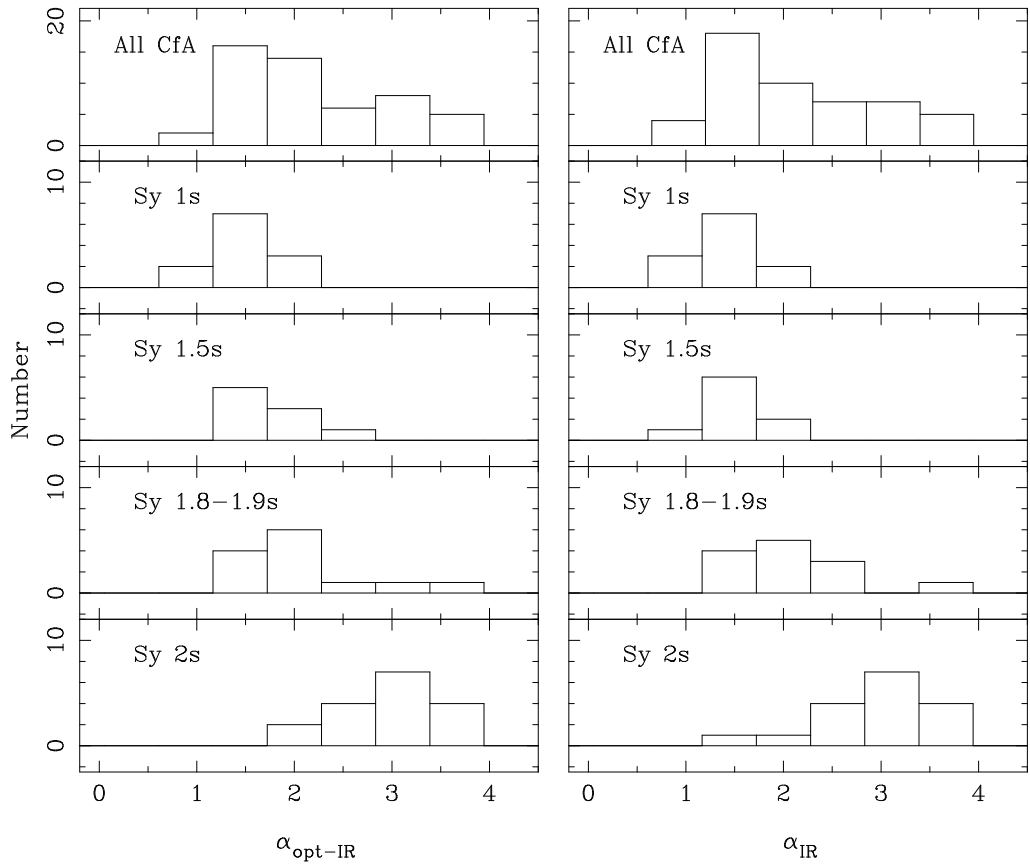}{425pt}{0}{80}{80}{-250}{50}
\vspace{-3cm}
\caption{Distributions of the
fitted spectral indices ($f_\nu \propto \nu^{-\alpha}$) 
in the optical-infrared wavelengths 
($\alpha_{\rm opt-IR}$ for the 
spectral range $\simeq 0.4-16\,\micron$,  left panels) 
and infrared wavelengths ($\alpha_{\rm IR}$ for the 
spectral range $1-16\,\micron$, right panels). 
The very top panels are the 51 Seyferts in the 'extended' 
CfA sample with measured spectral indices. The panels below are the same sample
broken up in different Seyfert types as derived 
from optical spectroscopy.}
\end{figure*}

\begin{figure*}
\figurenum{7}
\plotfiddle{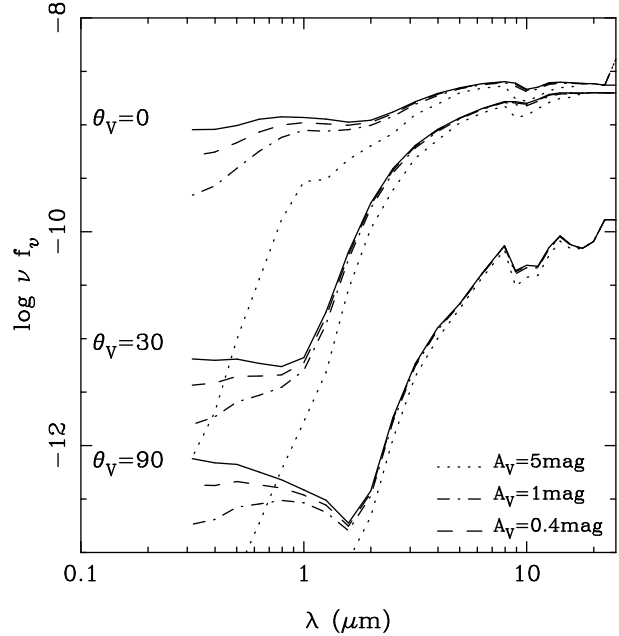}{425pt}{-90}{70}{70}{-150}{550}
\vspace{-5cm}
\caption{Effects of foreground extinction ($A_{\rm V} = 0.4, \, 1\,
{\rm and} \, 5\,$mag) on the observed SED of 
an AGN, using Rieke \& Lebofsky (1985) extinction 
law. The AGN SEDs are outputs of the Torus+Cone model 
from Efstathiou \&
Rowan-Robinson 1995 (see Section~6 and Table~8 for details) for three 
different viewing angles. 
$\theta_{\rm v} = 0\arcdeg$ is the polar view of the AGN, 
$\theta_{\rm v} = 30\arcdeg$ is an intermediate view (corresponding
to the half opening angle of the torus, $\Theta_{\rm c}$), and 
$\theta_{\rm v} = 90\arcdeg$ is the 
equatorial view (through the torus).
} 
\end{figure*}

\begin{figure*}
\figurenum{8}
\plotfiddle{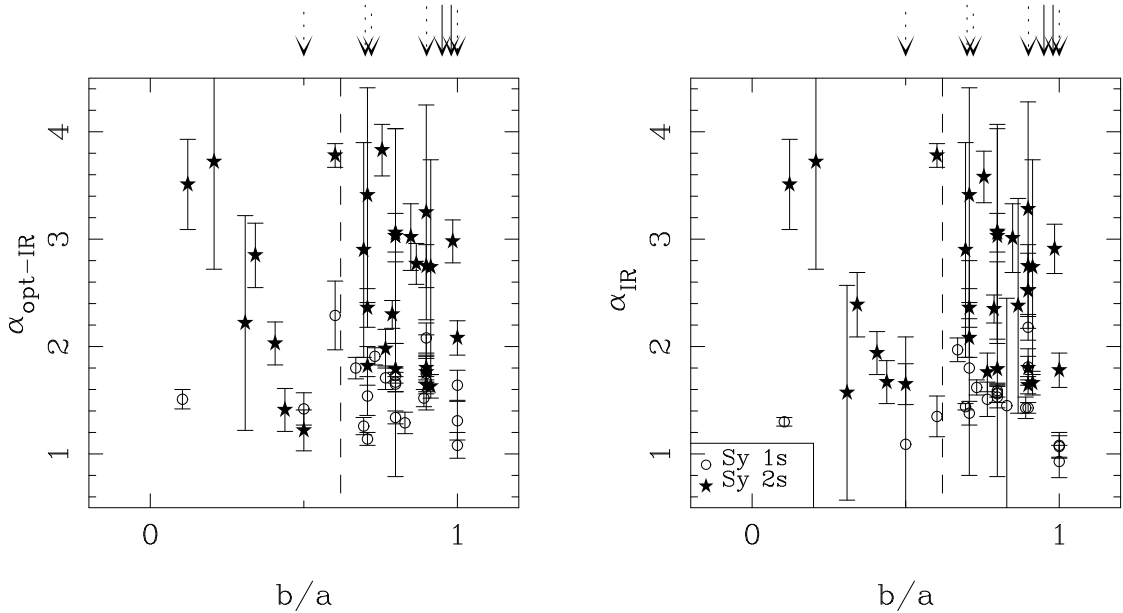}{425pt}{0}{90}{90}{-280}{160}
\vspace{-7cm}
\caption{Fitted optical-infrared (left panel) and 
infrared (right panel) spectral indices (see Section~5) 
as a function of the inclination
of the host galaxy, where $b/a=0$ is an edge-on galaxy, and 
$b/a=1$ is a face-on galaxy. The arrows at the top of the plots
indicate the values of $b/a$ for the remaining 7 galaxies with no measured 
spectral indices ---dotted lines for the type 2s (green 
in the on-line version) and solid lines for the type 
1s (red in the on-line version). The dashed line indicates a host galaxy 
axial ratio of $b/a=0.6$ or an inclination
of $i> 53\arcdeg$ (see discussion in text). The galaxies 
are separated between Seyfert 1s ($1-1.5$) shown as open circles (red in
the on-line version) and 
Seyfert 2s ($1.8-2$) shown 
as filled star symbols (green in the on-line version), 
using the  classification from optical spectroscopy.}
\end{figure*}

\begin{figure*}
\figurenum{9}
\plotfiddle{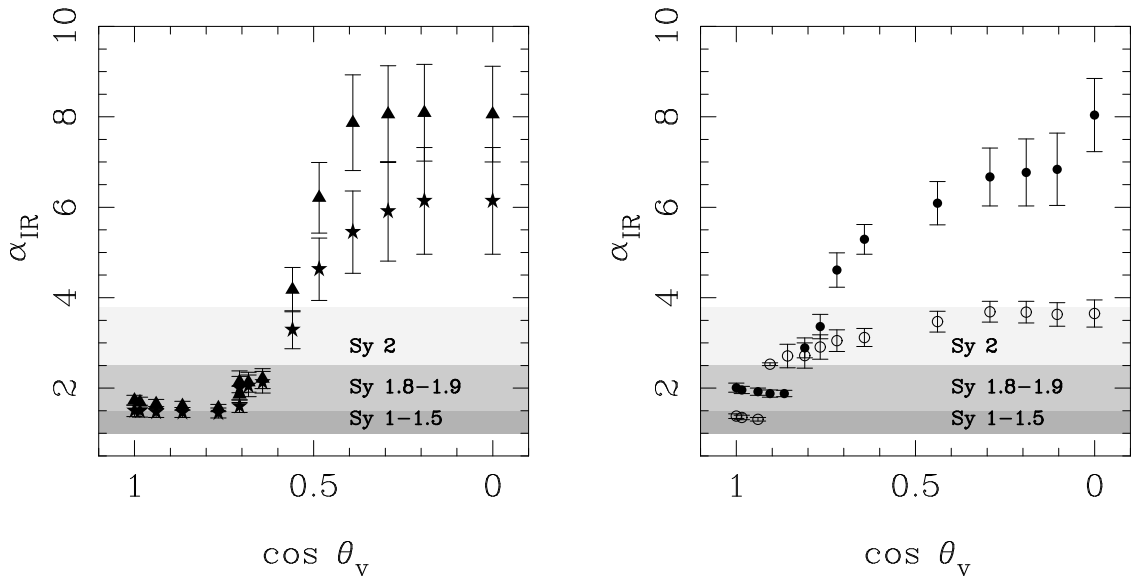}{425pt}{-90}{90}{90}{-280}{700}
\vspace{-6cm}
\caption{Predicted infrared spectral 
indices ($f_\nu \propto \nu^{-\alpha_{\rm IR}}$) 
in the $1-16\,\micron$ spectral 
range as a function of the cosine of the viewing angle to the torus 
($\theta_{\rm v} = 0\arcdeg$
polar view of the AGN, $\theta_{\rm v} = 90\arcdeg$
equatorial view of the AGN)  using 
Efstathiou \& Rowan-Robinson (1995) and 
Efstathiou et al. (1995) torus models. {\it Left panel:} 
results for models with a torus half opening angle
$\Theta_c =45\arcdeg$, and two different equatorial UV 
optical depths: $\tau_{\rm UV} = 500\,$mag (filled triangles),  
$\tau_{\rm UV} = 250\,$mag (filled stars), Models Torus 2 and Torus 3 
respectively (see Table~8). {\it Right panel:} results for 
a torus model with a half opening angle $\Theta_c =30\arcdeg$, and  
an equatorial UV optical depth of $\tau_{\rm UV} = 1200\,$mag
with the cone component (Model Torus+Cone in Table~8, open circles) and 
without the cone component
(Model Torus 4 in 
Table~8, filled circles). In both diagrams the shaded areas
represent the observed ranges of fitted $\alpha_{\rm IR}$ for
Seyfert $1-1.5$s (darkest shaded 
area), Seyferts $1.8-1.9$s (intermediate shaded area), 
and Seyfert 2s (lightest shaded area) in the CfA sample.
}
\end{figure*}

\end{document}